\documentclass[journal, 10pt]{IEEEtran}
\ifCLASSINFOpdf
\else
\fi

\usepackage{amsmath,amssymb,amsfonts,amstext}
\usepackage{bbm}
\usepackage{epsfig,overpic,psfrag}
\usepackage{graphicx}
\usepackage{algorithm}
\usepackage{algorithmic}
\usepackage{tikz}

\newtheorem{thm}{Theorem}
\newtheorem{lemma}{Lemma}

\begin{document}

\sloppy

\title{Fronthaul Compression and Transmit Beamforming Optimization for Multi-Antenna Uplink C-RAN}

\author{Yuhan Zhou,~\IEEEmembership{Member,~IEEE} and Wei Yu,~\IEEEmembership{Fellow,~IEEE}
\thanks{Copyright (c) 2015 IEEE. Personal use of this material is permitted. However, permission to use this material for any other purposes must be obtained from the IEEE by sending a request to pubs-permissions@ieee.org.

Manuscript received October 6, 2015; revised March 6, 2016; accepted
April 15, 2016. This work was supported in part by Huawei Technologies Canada Co., Ltd., and in part by the Natural Sciences and Engineering Research Council (NSERC) of Canada. This paper was presented in part at IEEE Globecom Workshop, Austin, Texas, USA, December 2014. The associate editor coordinating the review of this manuscript and approving it for publication was Dr. Paolo Banelli.

Y. Zhou was with the The Edward S. Rogers Sr. Department of Electrical
and Computer Engineering, University of Toronto, Toronto, ON M5S 3G4
Canada. He is now with Qualcomm Technologies Inc., San Diego, CA 92121
USA (email: yzhou@ece.utoronto.ca).

W. Yu is with The Edward S. Rogers Sr. Department of Electrical and
Computer Engineering, University of Toronto, Toronto, ON M5S 3G4
Canada (e-mail: weiyu@ece.utoronto.ca).
}}


\markboth{IEEE TRANSACTIONS ON SIGNAL PROCESSING,~VOL. 00,~NO. 0, XXX 2016}%
{Shell \MakeLowercase{\textit{et al.}}: Bare Demo of IEEEtran.cls for Journals}


\maketitle

\begin{abstract}
 This paper considers the joint fronthaul compression and transmit beamforming design for the uplink cloud radio access network (C-RAN), in which multi-antenna user terminals communicate with a cloud-computing based centralized processor (CP) through multi-antenna base-stations (BSs) serving as relay nodes. A compress-and-forward relaying strategy, named the virtual multiple-access channel (VMAC) scheme, is employed, in which the  BSs can either perform single-user compression or Wyner-Ziv coding to quantize the received signals and send the quantization bits to the CP via capacity-limited fronthaul links; the CP performs successive decoding with either successive interference cancellation (SIC) receiver or linear minimum-mean-square-error (MMSE) receiver. Under this setup, this paper investigates the joint optimization of the transmit beamformers at the users and the quantization noise covariance matrices at the BSs for maximizing the network utility. A novel weighted minimum-mean-square-error successive convex approximation (WMMSE-SCA) algorithm is first proposed for maximizing the weighted sum rate under the user transmit power and fronthaul capacity constraints with single-user compression. Assuming a heuristic decompression order, the proposed algorithm is then adapted for optimizing the transmit beamforming and fronthaul compression under Wyner-Ziv coding. This paper also proposes a low-complexity separate design consisting of optimizing transmit beamformers for the Gaussian vector multiple-access channel along with per-antenna quantizers with uniform quantization noise levels across the antennas at each BS. Numerical results show that with optimized beamforming and fronthaul compression, C-RAN can significantly improve the overall performance of conventional cellular networks. Majority of the performance gain comes from the implementation of SIC at the central receiver. Furthermore, the low complexity separate design already performs very close to the optimized joint design in regime of practical interest.
\end{abstract}



\section{Introduction}
To meet the exponentially increasing data demand in wireless communication driven by smartphones, tablets, and video streaming, modern cellular communication systems are moving towards densely deployed heterogenous networks consisting of base-stations (BSs) covering progressively smaller areas. As a consequence, inter-cell interference becomes the dominant performance limiting factor. Cloud radio access network (C-RAN) is a novel mobile network architecture that offers an efficient way for managing inter-cell interference~\cite{Simeone15}. In a C-RAN architecture, the baseband and higher-layers operations of the BSs are migrated to a cloud-computing based centralized processor (CP). By allowing coordination and joint signal processing across multiple cells, C-RAN provides a platform for implementing network multiple-input multiple-output (network MIMO), also known as coordinated multi-point (CoMP), which can achieve significantly higher data rates than conventional cellular networks~\cite{Gesbert10}.

\begin{figure} [t]
    \centering
    \begin{overpic}[width=0.46\textwidth]{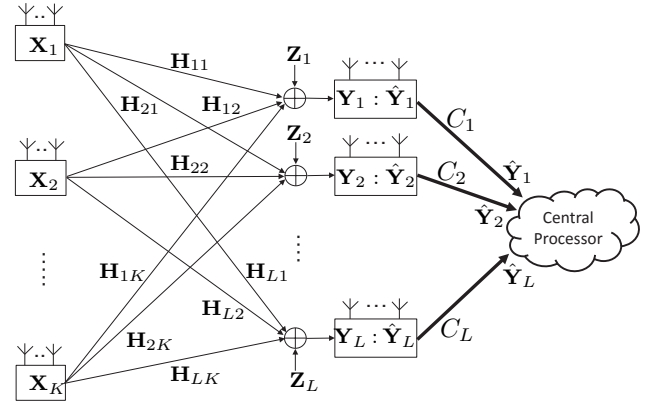}
    \put(51.5,47.5){\footnotesize $\mathbf{Y}_1:\hat{\mathbf{Y}}_1$}
    \put(51.5,35){\footnotesize $\mathbf{Y}_2:\hat{\mathbf{Y}}_2$}
    \put(51,9.5){\footnotesize $\mathbf{Y}_L\hspace{-.25mm}:\hspace{-.25mm}\hat{\mathbf{Y}}_L$}
    \put(2.5,56){\footnotesize $\mathbf{X}_1$}
    \put(2.5,34.5){\footnotesize $\mathbf{X}_2$}
    \put(2.5,2){\footnotesize $\mathbf{X}_K$}
    \put(43.5,54.5){\footnotesize $\mathbf{Z}_1$}
    \put(43.5,42){\footnotesize $\mathbf{Z}_2$}
    \put(44,2.5){\footnotesize $\mathbf{Z}_L$}
    \put(25,53.5){\footnotesize $\mathbf{H}_{11}$}
    \put(17,46.5){\footnotesize $\mathbf{H}_{21}$}
    \put(30,46.5){\footnotesize $\mathbf{H}_{12}$}
    \put(13.7,20){\footnotesize $\mathbf{H}_{1K}$}
    \put(37,20){\footnotesize $\mathbf{H}_{L1}$}
    \put(18,9){\footnotesize $\mathbf{H}_{2K}$}
    \put(30,14){\footnotesize $\mathbf{H}_{L2}$}
    \put(25,37.2){\footnotesize $\mathbf{H}_{22}$}
    \put(25,3.5){\footnotesize $\mathbf{H}_{LK}$}
    \put(78,35.5){\footnotesize $\hat{\mathbf{Y}}_1$}
    \put(73.5,28.5){\footnotesize $\hat{\mathbf{Y}}_2$}
    \put(78,19){\footnotesize $\hat{\mathbf{Y}}_L$}
    \put(69,44){$C_1$}
    \put(67.5,35.5){$C_2$}
    \put(68,10.5){$C_L$}
    \end{overpic}
    \caption{An uplink C-RAN system with capacity-limited fronthaul.}
    \label{fig:C-RAN}
\end{figure}

This paper focuses on the uplink C-RAN architecture as shown in Fig.~\ref{fig:C-RAN}, where multi-antenna mobile users communicate with the CP with multi-antenna BSs serving as relay nodes. The BSs are connected with the CP via digital fronthaul links with finite capacities. We consider a two-stage compress-and-forward relaying strategy, referred to as the virtual multiple access channel (VMAC) scheme, in which the BSs quantize the received signals using either single-user compression or Wyner-Ziv coding and send the compressed bits to the CP. The CP performs successive decoding to decode the quantization codewords first, then the user messages sequentially. Under the VMAC scheme, this paper studies the optimization of the transmit beamforming vectors and quantization noise covariance matrices for maximizing the weighted sum rate of the C-RAN system. Being different from the conventional multicell cellular systems, in which the optimal transmit beamforming only depends on the interfering signal strength and the channel gain matrices, in C-RAN, the finite fronthaul capacity also needs to be taken into account in the beamforming design.
This paper proposes a novel weighted minimum-mean-square-error successive convex approximation (WMMSE-SCA) algorithm to find the optimal transmit beamformers and quantization noise covariance matrices for maximizing the weighted sum rate of C-RAN. Moreover, a simple separate design consisting of optimizing transmit beamformers for the Gaussian vector multiple-access channel and per-antenna quantizers with uniform quantization noise levels across the antennas at each BS is also developed, under the assumption that the signal-to-quantization-noise ratio (SQNR) is high and successive interference cancellation (SIC) is applied at the receiver. Numerical simulations show that the proposed separate design already performs very close to the optimized joint design in the SQNR regime of practical interest.

This paper considers two different fronthaul compression strategies for C-RAN, namely \emph{single-user compression} and \emph{Wyner-Ziv coding}. In single-user compression, which is also referred to as point-to-point compression in the literature~\cite{park2014fronthaul}, each BS uses vector quantization to compress the received signals but ignores the correlation between the received signals across different BSs. In contrast, Wyner-Ziv coding fully utilizes the correlation of the received signals for higher compression efficiency, thereby achieving better overall performance. The optimization strategy proposed in this paper is first developed for single-user compression, then for the more complex Wyner-Ziv coding, assuming a heuristic ordering for decompression of the quantized signals at the BSs.
The performance of the VMAC schemes with single-user compression and Wyner-Ziv coding are evaluated for practical multicell networks under linear \emph{minimum-mean-square-error (MMSE) receiver} and  \emph{SIC receiver} respectively. It is shown that the implementation of SIC receiver significantly improves the performance achieved by linear MMSE receiver under both single-user compression and Wyner-Ziv coding. Furthermore, although single-user compression with SIC receiver can already realize majority of the benefit brought by the C-RAN architecture, Wyner-Ziv coding can further improve upon single-user compression when the fronthaul capacity is limited.

To precisely quantify the advantage of the C-RAN architecture, this paper further evaluates the performance of optimized beamforming and fronthaul compression under two different types of BS clustering strategies: \emph{disjoint clustering} and  \emph{user-centric clustering}. In disjoint clustering scheme, the entire network is divided into non-overlapping clusters and the BSs in each cluster jointly serve all the users within the coverage area~\cite{3GPP11, peng2015fronthaul}. In user-centric clustering, each user is served by an individually selected subset of neighboring BSs;  different clusters for different users may overlap.
The performance of user-centric clustering has been evaluated for the downlink of cooperative cellular networks~\cite{ng2010linear} and C-RAN systems~\cite{dai2014sparse}. This paper further shows numerically that in uplink C-RAN, with optimized beamforming and fronthaul compression, the user-centric clustering strategy significantly outperforms the disjoint clustering strategy, because the cell edges are effectively eliminated.

\subsection{Related Work}

One of the main issues in the implementation of C-RAN is how to optimally utilize the capacity-limited fronthaul links to efficiently reap the benefit of multicell processing. Substantial research works have made progress towards this direction~\cite{Sand08, zhoulei2013, Zhou14Globecom}. Under the assumption of compress-and-forward relaying strategy at the the BSs, the largest achievable rate for uplink C-RAN is given by the joint decompression and decoding strategy, in which quantization codewords and user messages are decoded simultaneously~\cite{Sand09, Lim11, Park13}. However, the complexity of joint decompression and decoding is very high, which prevents it from practical implementation. In~\cite{zhou2014JSAC}, a virtual multiple access channel (VMAC) scheme, which is a compress-and-forward strategy based on the successive decoding of quantization codewords followed by user messages, is proposed for the single-input single-output (SISO) C-RAN architecture. As compared to the joint decompression and decoding scheme, the VMAC scheme has lower decoding complexity and shorter decoding delay, which makes it more desirable for practical implementation. Furthermore, it is shown in~\cite{zhou2015ISIT} that with Wyner-Ziv coding the successive decoding based VMAC scheme actually achieves the same maximum sum rate as the joint decompression and decoding strategy for the uplink C-RAN model under a sum fronthaul constraint.

This paper studies the linear transceiver and fronthaul compression design in the VMAC scheme for the uplink multiple-input-multiple-output (MIMO) C-RAN model. As a generalization of \cite{zhou2014JSAC} which considers the SISO case only, this paper considers the MIMO case where both the users and the BSs are equipped with multiple antennas. The main difference between the SISO case and the MIMO case is the impact of transmitter optimization at the user terminals. For example, in the SISO case, to maximize the sum rate of the VMAC scheme for a single-cluster C-RAN system, all the users within the cluster need to transmit at their maximum powers (assuming that SIC receiver is implemented at the CP). However, in the MIMO case, the users are capable of doing transmit beamforming, so the optimal transmit beamforming design is more involved.

The fronthaul compression problem for the uplink C-RAN model has been considered extensively
in the literature. Various algorithms such as alternating convex optimization~\cite{zhou2014JSAC}, gradient projection~\cite{Del09}, and the robust fronthaul compression approach~\cite{Park12} have been developed for maximizing the (weighted) sum rate under the fronthaul constraints. All of these algorithms focus only on the optimization of quantization noise covariance matrices across the BSs, with fixed transmit beamformers. This paper goes one step further by considering the joint transmit beamformer and quantization noise covariance matrix optimization problem. Accounting for both the transmit beamforming and the quantization design problem together in the optimization framework is nontrivial, because the two are coupled through the fronthaul constraints. To tackle this problem, this paper proposes a novel WMMSE-SCA algorithm for efficiently finding a local optimum solution to the weighted sum rate maximization problem. The proposed algorithm integrates the well-known WMMSE beamforming design strategy~\cite{christensen2008weighted, shi2011iteratively}, with  the successive convex approximation technique~\cite{hunter2004tutorial, razaviyayn2013unified}, to arrive at a stationary point of the maximization problem. The performance of optimized beamforing vectors and quantization noise covariance matrices for both Wyner-Ziv coding and single-user compression are evaluated under practical multicell networks with different receive beamforming schemes, i.e., the linear MMSE receiver and the SIC receiver. Simulation results show that the performance improvement of the SIC receiver as compared to the linear MMSE receiver is much larger than that of Wyner-Ziv coding as compared to single-user compression. Most of the performance gain brought by C-RAN can thus be obtained by single-user compression together with SIC receiver.

From a broader perspective, this paper studies the radio resource allocation optimization for uplink fronthaul-constrained C-RAN~\cite{peng2015fronthaul}. As related work, we mention \cite{rao2015distributed} which proposes to utilize the signal sparsity in C-RAN to improve the performance of the fronthaul compression and user detection. The fronthaul compression can also be designed for enhancing synchronization in C-RAN~\cite{heo2015optimal}. Additionally, we mention briefly that the latency issue in the C-RAN design has been studied in~\cite{wang2015delay}, which introduces a delay-optimal fronthaul allocation strategy for the latency control. Finally, we point out that with proper modification the design idea of fronthaul-aware beamforming proposed in this paper can be extended to the more recently proposed heterogeneous C-RAN~\cite{peng2014heterogeneous} and Fog RAN~\cite{peng2015fog} architectures.

\subsection{Paper Organization and Notation}
The rest of the paper is organized as follows. Section \ref{sec:system}
introduces the system model and the VMAC scheme.  Section \ref{sec:joint-design-SU}
considers the joint design of beamforming and fronthaul compression under single-user compression,
where a novel WMMSE based successive convex optimization algorithm is proposed.
The proposed joint design scheme is developed further in Section \ref{sec:joint-design-WZ} for maximizing weighted sum rate under Wyner-Ziv coding.
Section \ref{sec:separate-design} is devoted to a low-complexity separate beamforming and fronthaul compression design, which is shown to be near-optimal at high SQNR regime.
The proposed algorithms are evaluated numerically for practical multicell and multicluster networks in
Section \ref{sec:simu}. Conclusions are drawn in Section \ref{sec:conclusion}.

The notations used in this paper are as follows. Boldface lower-case letters denote column vectors. Boldface upper-case letters denote vector random variables or matrices, where
context should make the distinction clear. Superscripts $(\cdot)^T$,
$(\cdot)^{\dag}$, and $(\cdot)^{-1}$ denote transpose,
Hermitian transpose, and matrix inverse operators; $\mathbb{E}(\cdot)$, $\|\cdot\|_{F}$, $\mathrm{Tr}(\cdot)$, and $\mathrm{rank}(\cdot)$ denote expectation, Frobenius norm, matrix trace, and matrix rank operators.
For a vector/matrix $\mathbf{X}$, $\mathbf{X}_{\mathcal{S}}$ denotes a vector/matrix with
elements whose indices are elements of $\mathcal{S}$. Given matrices $\{\mathbf{X}_1,\ldots, \mathbf{X}_L\}$, $\mathrm{diag}\left(\{\mathbf{X}_{\ell}\}_{\ell=1}^L\right)$ denotes
the block diagonal matrix formed with $\mathbf{X}_{\ell}$ on the diagonal. We let $\mathcal{K}=\{1,\cdots,K\}$ and $\mathcal{L}=\{1,\cdots,L\}$.

\section{Preliminaries}
\label{sec:system}
\subsection{System Model}
This paper considers the uplink C-RAN, where $K$ multi-antenna mobile users communicate with
a CP through $L$ multi-antenna BSs serving as relay nodes, as shown in Fig.~\ref{fig:C-RAN}. The noiseless fronthaul
links connecting the BSs with the CP have per-link capacity of $C_{\ell}$. Each user
terminal is equipped with $M$ antennas; each BS is equipped with $N$ antennas.

We consider the VMAC scheme~\cite{zhou2014JSAC} applied to such a C-RAN system, in which the BSs quantize the received signals
using either Wyner-Ziv coding or single-user compression, then forward the compressed bits to the CP for decoding.
In single-user compression, the compression process only involves the conventional vector quantizers, one for each BS, while in Wyner-Ziv coding,
the correlation between the received signals across the BSs are fully utilized for higher compression efficiency.
At the CP side, a two-stage successive decoding strategy is employed, where the quantization codewords are first decoded,
and then the user messages are decoded sequentially.

We assume that the wireless channel between the users and the BSs are quasi-static, with large coherence
time/bandwidth. Furthermore, perfect channel state information (CSI)
is assumed to be available at the BSs. The BSs forward the CSI to the CP through fronthaul links. The CP determines the optimal quantization noise covariance matrices, then feeds them back to the BSs. The CP also determines the user transmit beamformers and feeds them back to the users.
Under the quasi-static channel, the overheads due to the CSI transmission and feedback between the BSs and the CP can be amortized
within the channel coherence time and is ignored in this paper for simplicity.

Define $\mathbf{H}_{\ell, k}$ as the $N\times M$ complex channel matrix
between the $k$th user and the $\ell$th BS.
It is assumed that each user intends to transmit $d$ parallel data streams to the CP. (Throughout this paper we assume that $d$ is fixed.) Let $\mathbf{V}_k \in \mathbb{C}^{M\times d}$ denote the linear transmit beamfomer that user $k$ utilizes to transmit message signal $\mathbf{s}_k \in \mathbb{C}^{d\times 1}$ to the central receiver. We assume that each message signal $\mathbf{s}_k$ intended for user $k$ is taken from a Gaussian codebook so that we have $\mathbf{s}_k\thicksim \mathcal{CN}(\mathbf{0}, \mathbf{I})$. Then the transmit signal at user $k$ is given by $\mathbf{x}_k = \mathbf{V}_k \mathbf{s}_k$ with $\mathbb{E}[\mathbf{x}_k\mathbf{x}_k^{\dag}]=\mathbf{V}_k \mathbf{V}_k^{\dag}$. The transmit beamformers are subjected to per-user power constraints, i.e., $\mathrm{Tr}\left(\mathbf{V}_k \mathbf{V}_k^{\dag}\right) \leq P_k$ for $k\in \mathcal{K}$. The received signal at BS $\ell$, $\mathbf{y}_{\ell}$, can be expressed as
\begin{equation}
\mathbf{y}_{\ell}  = \sum_{k=1}^K \mathbf{H}_{\ell, k} \mathbf{V}_k \mathbf{s}_k + \mathbf{z}_{\ell},  \quad \forall \;  \ell \in \mathcal{L},
\label{eqn:received-sig-Y}
\end{equation}
where $\mathbf{z}_{\ell}\sim\mathcal{CN}(\mathbf{0},\mathbf{\Lambda}_{\ell})$ represents the additive Gaussian noise for BS $\ell$.
Assuming Gaussian quantization test channel, the quantized received signal $\mathbf{\hat{y}}_{\ell}$ for the $\ell$th BS is given by
\begin{equation}
\mathbf{\hat{y}}_{\ell} = \mathbf{y}_{\ell} + \mathbf{q}_{\ell}
\label{eqn:quantized-Y}
\end{equation}
where $\mathbf{q}_{\ell}\thicksim \mathcal{CN}(\mathbf{0}, \mathbf{Q}_{\ell})$ represents the Gaussian quantization noise for the $\ell$th BS.

The above Gaussian quantization test channel model (\ref{eqn:quantized-Y})
is sufficiently general to encompass the possibility of performing
receive beamforming at the BSs prior to quantization.
For a given quantization noise covariance matrix $\mathbf{Q}_{\ell}$, let
$\mathbf{Q}_{\ell} = \mathbf{B}_{\ell} \mathbf{\Phi}_{\ell} \mathbf{B}_{\ell}^{\dagger}$
be its eigenvalue composition, 
where $\mathbf{\Phi}_{\ell} = \mathrm{diag}\left(\left\{\varphi^i_{\ell}\right\}_{i=1}^N\right)$ with
$\varphi^i_{\ell}$ as the $i$th eigenvalue of $\mathbf{Q}_{\ell}$, and
$\mathbf{B}_{\ell}$ is the $N\times N$ unitary matrix whose $i$th column is
the eigenvector $\mathbf{b}^i_{\ell}$ corresponding to eigenvalue
$\varphi^i_{\ell}$. Then, setting quantization noise covariance to be
$\mathbf{Q}_{\ell}$ is equivalent to beamforming with $\mathbf{B}^{\dagger}_{\ell}$
followed by independent quantization, i.e.,
\begin{equation}
\mathbf{\hat{y}'}_{\ell} =
\mathbf{B}^{\dagger}_{\ell} \mathbf{y}_{\ell} + \tilde{\mathbf{q}}_{\ell}
\end{equation}
where $\mathbf{B}_{\ell}= \left[\mathbf{b}^1_{\ell},\ldots,\mathbf{b}^N_{\ell}\right]$ is the receive beamforming at the $\ell$th BS, and $\tilde{\mathbf{q}}_{\ell} \thicksim \mathcal{CN}(\mathbf{0}, \mathbf{\Phi}_{\ell})$ represents the quantization noise after the beamforming.

\subsection{Achievable Rate of the VMAC scheme}
The rate region of the VMAC scheme is characterized by that of a multiple-access channel, in which multiple users send information to a common CP. Following the results in \cite{zhou2014JSAC}, assuming that the linear MMSE receiver is applied at the CP, the transmission rate $R_k$ for user $k$ for the VMAC scheme is given by
\begin{eqnarray}\label{eqn:rate-express}
R_k &\leq& I(\mathbf{X}_k; \mathbf{Y}_1,\ldots, \mathbf{Y}_L) \nonumber \\
&=& \log\left|\mathbf{I} + \mathbf{V}_k^{\dag} \mathbf{H}^{\dag}_{\mathcal{L},k}\mathbf{J}_k^{-1}\mathbf{H}_{\mathcal{L},k} \mathbf{V}_k \right|
\end{eqnarray}
where
\begin{equation}\label{eqn:J}
\mathbf{J}_k = \mathbf{J}^{\mathrm{LE}}_k = \sum_{j \neq k}^K \mathbf{H}_{\mathcal{L},j} \mathbf{V}_j \mathbf{V}_j^{\dag}\mathbf{H}^{\dag}_{\mathcal{L},j} +  \mathbf{\Lambda} + \mathbf{Q},
\end{equation}
with $\mathbf{\Lambda} = \mathrm{diag}\left(\{\mathbf{\Lambda}_{\ell}\}_{\ell=1}^L\right)$ and $\mathbf{Q} = \mathrm{diag}\left(\{\mathbf{Q}_{\ell}\}_{\ell=1}^L\right)$. To achieve higher throughput, the SIC scheme can also be combined with the linear MMSE receiver. In this case, assuming without loss of generality a decoding order of user messages $1,2,\ldots,K$,
the matrix $\mathbf{J}_k = \mathbf{J}^{\mathrm{LE}}_k$ in (\ref{eqn:rate-express}) is replaced by $\mathbf{J}^{\mathrm{SIC}}_k$ expressed as
\begin{equation}\label{eqn:J-SIC}
\mathbf{J}_k = \mathbf{J}^{\mathrm{SIC}}_k = \sum_{j > k}^K \mathbf{H}_{\mathcal{L},j} \mathbf{V}_j \mathbf{V}_j^{\dag}\mathbf{H}^{\dag}_{\mathcal{L},j} + \mathbf{\Lambda} + \mathbf{Q}.
\end{equation}

The compression rates at the BSs should also satisfy the fronthaul link capacity constraints. Using information-theoretic formulation, the fronthaul constraints under single-user compression can be written as
\begin{equation}
\label{eqn:fronthaul-I-SU}
I\left(\mathbf{Y}_{\mathcal{\ell}}; \hat{\mathbf{Y}}_{\mathcal{\ell}}\right) \leq C_{\ell}, \;\forall \; \ell \in \mathcal{L}.
\end{equation}
Evaluating the above mutual information expression with Gaussian input and Gaussian quantization noise, the fronthaul constraint (\ref{eqn:fronthaul-I-SU}) becomes~\cite{zhou2013approximate}
\begin{equation}
\label{eqn:fronthaul-contraint-SU}
\log\frac{\left|\sum_{k=1}^K\mathbf{H}_{\ell, k} \mathbf{V}_k \mathbf{V}_k^{\dag}\mathbf{H}_{\ell,k}^{\dag} + \mathbf{\Lambda}_{\ell} + \mathbf{Q}_{\ell}\right|}{\left|\mathbf{Q}_{\ell}\right|} \leq C_{\ell}, 
\end{equation}
for all $\ell = 1,2,\ldots,L$.
When Wyner-Ziv coding is implemented at BSs, the fronthaul constraints are given by the following mutual information expressions~\cite{zhou2015ISIT,zhou2013ITW}
\begin{equation}
\label{eqn:fronthaul-I-WZ}
I\left(\mathbf{Y}_{\mathcal{S}}; \hat{\mathbf{Y}}_{\mathcal{S}}| \hat{\mathbf{Y}}_{\mathcal{S}^c}\right) \leq \sum_{\ell\in \mathcal{S}}C_{\ell}, \quad \forall \; \mathcal{S} \subseteq \mathcal{L}.
\end{equation}
Utilizing the chain rule on mutual information and the  Gaussian assumption, one can express the fronthaul constraint (\ref{eqn:fronthaul-I-WZ}) for Wyner-Ziv coding as follows,
\begin{multline}
\label{eqn:fronthaul-contraint-WZ}
\displaystyle \log\frac{\left|\sum\limits_{k=1}^K\mathbf{H}_{\mathcal{L},k} \mathbf{V}_k \mathbf{V}_k^{\dag}
\mathbf{H}^{\dagger}_{\mathcal{L},k}  +
\mathrm{diag}\left(\{\mathbf{\Lambda}_{\ell} +
\mathbf{Q}_{\ell}\}_{\ell \in \mathcal{L}}\right)\right|}{\left|\sum\limits_{k=1}^K\mathbf{H}_{\mathcal{S}^c,k} \mathbf{V}_k \mathbf{V}_k^{\dag}
\mathbf{H}^{\dagger}_{\mathcal{S}^c,k}  +
\mathrm{diag}\left(\{\mathbf{\Lambda}_{\ell} +
\mathbf{Q}_{\ell}\}_{\ell \in \mathcal{S}^c}\right) \right|} \\
 - \sum_{\ell\in \mathcal{S}}\log| \mathbf{Q}_{\ell}| \leq \sum_{\ell\in \mathcal{S}}C_{\ell},
\quad \forall \; \mathcal{S}\subseteq\mathcal{L}.
\end{multline}


\section{Joint Beamforming and Compression Design under Single-User Compression}
\label{sec:joint-design-SU}
\subsection{Weighted Sum Rate Maximization}
This section investigates the joint beamforming and fronthaul compression design for the VMAC scheme with single-user compression. As shown in the achievable rate expression (\ref{eqn:rate-express}) and the fronthaul constraint expression (\ref{eqn:fronthaul-contraint-SU}), the beamforming vectors and quantization noise covariance matrices are coupled, and the two together determine the overall performance of a C-RAN system. To characterize the tradeoff between the achievable rates for the users and the system resources, we formulate the following weighted sum rate maximization problem:
\begin{subequations} \label{prob:weight-sumrate-SU}
\begin{eqnarray}
\hspace{-10mm}& \displaystyle \max_{\mathbf{V}_k,\mathbf{Q}_{\ell}} \label{wsr-obj} &
\sum_{k=1}^K \alpha_k \log\left|\mathbf{I} + \mathbf{V}_k^{\dag} \mathbf{H}^{\dag}_{\mathcal{L},k}\mathbf{J}_k^{-1}\mathbf{H}_{\mathcal{L},k} \mathbf{V}_k \right| \\
\hspace{-10mm}& \mathrm{s.t.} &  \mathbf{J}_k = \mathbf{J}^{\mathrm{LE}}_k \quad \mathrm{or} \quad \mathbf{J}_k = \mathbf{J}^{\mathrm{SIC}}_k, \nonumber \\
\hspace{-10mm}& & 
 \log\frac{\left|\sum_{k=1}^K\mathbf{H}_{\ell, k} \mathbf{V}_k \mathbf{V}_k^{\dag}\mathbf{H}_{\ell,k}^{\dag} + \mathbf{\Lambda}_{\ell} + \mathbf{Q}_{\ell}\right|}{\left|\mathbf{Q}_{\ell}\right|} \leq C_{\ell}, \; \label{wsr-fhaulconst} \\
\hspace{-10mm} & & \mathbf{Q}_{\ell} \succeq \mathbf{0}, \quad \forall \ell \in \mathcal{L}, \nonumber \\
\hspace{-10mm}& & \mathrm{Tr}\left(\mathbf{V}_k \mathbf{V}_k^{\dag}\right) \leq P_k, \quad \forall k \in \mathcal{K} \nonumber
\end{eqnarray}
\end{subequations}
where $\alpha_k$'s are the weights representing the priorities associated
with the mobile users typically determined from upper layer protocols. When SIC receiver is implemented, to maximize the weighed sum rate, the user with larger weight should be decoded last. Without loss of generality, we assume $0 \leq \alpha_1\leq  \alpha_2 \leq\cdots\leq\alpha_K$, which results in the decoding order of user messages $1,2,\ldots,K$.

Due to the non-convexity of both the objective function and the fronthaul capacity constraints in problem (\ref{prob:weight-sumrate-SU}), finding the global optimum solution of (\ref{prob:weight-sumrate-SU}) is challenging. We point out here that the present formulation (\ref{prob:weight-sumrate-SU}) actually implicitly includes the user scheduling strategy. More specifically, one can consider a weighted sum rate maximization problem over all the users in the network, where the beamformers for the users are set to be the zero vector if they are not scheduled. For simplicity in the following development, we focus on the active users only and assume that user scheduling is done prior to solving problem (\ref{prob:weight-sumrate-SU}). Implicit scheduling is discussed later in the simulation part of the paper.

\subsection{The WMMSE-SCA Algorithm}
In this section, we propose a novel algorithm to find a stationary point of the problem (\ref{prob:weight-sumrate-SU}).
The main difficulty in solving (\ref{prob:weight-sumrate-SU}) comes from
the fact that the objective function and fronthaul capacity constraints are both nonconvex functions with respect to the optimization variables.
Inspired by the recent work of using the WMMSE approach for beamforming design \cite{christensen2008weighted, shi2011iteratively},
we first reformulate the objective function in problem (\ref{prob:weight-sumrate-SU}) as a convex function with respect to the MMSE matrix given by the user's target signal $s_k$ and decoded signal $\hat{s}_k$.
We then linearize the convex objective function and the compression rate expressions in the fronthaul constraints of (\ref{prob:weight-sumrate-SU}) to obtain a convex approximation of the original problem. Finally we
successively approximate the optimal solution by optimizing
this convex approximation. The idea of convex approximation is rooted from modern optimization techniques including
block successive minimization method 
and minorize-maximization algorithm, which have been
previously applied for solving related problems in wireless communications~\cite{ng2010linear, park2013joint}.

Before presenting the proposed algorithm, we first state the following lemma, which is a direct consequence of concavity of the $\log |\cdot|$ function.

\begin{lemma}
\label{lem:logdeter-lowerbound}
For positive definite Hermitian matrices $\mathbf{\Omega}, \mathbf{\Sigma} \in \mathbb{C}^{N\times N}$,
\begin{equation}\label{eqn:logdeter-lowerbound}
\log \left|\mathbf{\Omega}\right| \leq \log\left|\mathbf{\Sigma}\right| + \mathrm{Tr}\left(\mathbf{\Sigma}^{-1}\mathbf{\Omega}\right)-N
\end{equation}
with equality if and only if $\mathbf{\Omega} = \mathbf{\Sigma}$.
\end{lemma}

By applying Lemma \ref{lem:logdeter-lowerbound} to the first log-determinant term in the fronthaul constraint expression (\ref{eqn:fronthaul-contraint-SU}) or (\ref{wsr-fhaulconst}) and by setting
\begin{equation}
\label{eqn:omga}
\mathbf{\Omega}_{\ell} = \sum_{k=1}^K\mathbf{H}_{\ell, k} \mathbf{V}_k \mathbf{V}_k^{\dag}\mathbf{H}_{\ell,k}^{\dag} + \mathbf{\Lambda}_{\ell} + \mathbf{Q}_{\ell},
\end{equation}
we can approximate the fronthaul constraint (\ref{eqn:fronthaul-contraint-SU}) or (\ref{wsr-fhaulconst}) with the following convex constraint:
\begin{multline}\label{eqn:const-linear}
 \log\left|\mathbf{\Sigma}_{\ell}\right| + \mathrm{Tr}\left(\mathbf{\Sigma}_{\ell}^{-1}\left(\sum_{k=1}^K\mathbf{H}_{\ell, k} \mathbf{V}_k \mathbf{V}_k^{\dag}\mathbf{H}_{\ell,k}^{\dag} + \mathbf{\Lambda}_{\ell} + \mathbf{Q}_{\ell}\right)\right) \\
 - \log\left|\mathbf{Q}_{\ell}\right| \leq C_{\ell} + N
\end{multline}
for $\ell =1,2,\ldots,L$. It is not hard to see that the fronthaul constraint (\ref{eqn:fronthaul-contraint-SU}) or (\ref{wsr-fhaulconst}) is always feasible when the convex constraint (\ref{eqn:const-linear}) is feasible. The two constraints are equivalent when
\begin{equation}
\label{eqn:opt-sigma}
\mathbf{\Sigma}^*_{\ell} = \sum_{k=1}^K\mathbf{H}_{\ell, k} \mathbf{V}_k \mathbf{V}_k^{\dag}\mathbf{H}_{\ell,k}^{\dag} + \mathbf{\Lambda}_{\ell} + \mathbf{Q}_{\ell}.
\end{equation}

Now we approximate the objective function (\ref{wsr-obj}) using the WMMSE approximation. Let $\mathbf{U}_k \in \mathbb{C}^{NL\times d}$ be the linear receiver applied at the CP for recovering $\mathbf{s}_k$. The transmission rate $R_k$ in (\ref{eqn:rate-express}) can be expressed as the following~\cite{christensen2008weighted}~\cite{shi2011iteratively},
\begin{equation}\label{eqn:rate-MMSE}
R_k = \max_{\mathbf{U}_k} \log |\mathbf{E}^{-1}_k|
\end{equation}
where
\begin{multline}\label{eqn:E-matrix}
\mathbf{E}_k = (\mathbf{I} - \mathbf{U}_k^{\dag}\mathbf{H}_{\mathcal{L},k}\mathbf{V}_k)(\mathbf{I} - \mathbf{U}_k^{\dag}\mathbf{H}_{\mathcal{L},k}\mathbf{V}_k)^{\dag} \\
+ \mathbf{U}_k^{\dag}\left(\sum_{j\neq k}^K \mathbf{H}_{\mathcal{L},j} \mathbf{V}_j \mathbf{V}_j^{\dag}\mathbf{H}^{\dag}_{\mathcal{L},j} + \mathbf{\Lambda} + \mathbf{Q}\right)\mathbf{U}_k.
\end{multline}

By applying Lemma \ref{lem:logdeter-lowerbound} again, we rewrite rate expression (\ref{eqn:rate-MMSE}) as
\begin{equation}\label{eqn:rate-WMMSE}
R_k = \max_{\mathbf{W}_k, \mathbf{U}_k}\left(\log |\mathbf{W}_k| - \mathrm{Tr}(\mathbf{W}_k\mathbf{E}_k) + d\right)
\end{equation}
where $\mathbf{W}_k$ is the weight matrix introduced by the WMMSE method. The optimal $\mathbf{W}_k$ is given by
\begin{equation}
\label{eqn:opt-weightmatrix}
\mathbf{W}^*_k = \mathbf{E}^{-1}_k = \mathbf{I} +  \mathbf{V}^{\dag}_k\mathbf{H}^{\dag}_{\mathcal{L},k}\mathbf{U}^*_{k},
\end{equation}
where $\mathbf{U}^*_{k}$ is the MMSE receive beamformer given by
\begin{equation}
\label{eqn:opt-receiver}
\mathbf{U}^*_k = \left(\sum_{j\neq k}\mathbf{H}_{\mathcal{L}, j} \mathbf{V}_j \mathbf{V}_j^{\dag}\mathbf{H}_{\mathcal{L},j}^{\dag} + \mathbf{\Lambda} + \mathbf{Q}\right)^{-1}\mathbf{H}_{\mathcal{L},k} \mathbf{V}_k.
\end{equation}

Using (\ref{eqn:rate-WMMSE}) and (\ref{eqn:const-linear}) to replace the objective function and the fronthaul constraints in problem (\ref{prob:weight-sumrate-SU}), we reformulate the weighted sum-rate maximization problem as follows
\begin{eqnarray}\label{prob:wsumrate-reform}
\hspace{-5mm}& \displaystyle \max_{\substack{\mathbf{V}_k, \mathbf{Q}_{\ell}, \mathbf{U}_k, \\ \mathbf{W}_k,\mathbf{\Sigma}_{\ell},\mathbf{\Theta}_{\ell}}} &  \sum_{k=1}^K \alpha_k \left( \log |\mathbf{W}_k| - \mathrm{Tr}(\mathbf{W}_k \mathbf{E}_k) \right)\nonumber \\
\hspace{-5mm}& & \hspace{2cm} - \rho\sum_{\ell=1}^L\left\|\mathbf{Q}_{\ell} -\mathbf{\Theta}_{\ell}\right\|^2_{F} \\
\hspace{-5mm}& \mathrm{s.t.} & \log\left|\mathbf{\Sigma}_{\ell}\right| + \mathrm{Tr}\left(\mathbf{\Sigma}_{\ell}^{-1}\mathbf{\Omega}_{\ell}\right) - \log\left|\mathbf{Q}_{\ell}\right| \leq C'_{\ell},  \;  \forall \ell \in \mathcal{L}, \nonumber\\
\hspace{-5mm}& & \mathbf{Q}_{\ell} \succeq \mathbf{0}, \enspace \mathbf{\Theta}_{\ell} \succeq \mathbf{0}, \quad \forall \ell \in \mathcal{L}, \nonumber \\
\hspace{-5mm}& & \mathrm{Tr}\left(\mathbf{V}_k \mathbf{V}_k^{\dag}\right) \leq P_k, \quad \forall k \in \mathcal{K},  \nonumber
\end{eqnarray}
where $\mathbf{\Omega}_{\ell} = \sum_{k=1}^K\mathbf{H}_{\ell, k} \mathbf{V}_k \mathbf{V}_k^{\dag}\mathbf{H}_{\ell,k}^{\dag} + \mathbf{\Lambda}_{\ell} + \mathbf{Q}_{\ell}$, $\rho$ is some positive constant, and $C'_{\ell} = C_{\ell} + N$. Note that the last term in the objective function which involves a summation of Frobenius norms is a quadratic regularization term. It makes the optimization problem (\ref{prob:wsumrate-reform}) strictly convex with respect to each optimization variable.

It is easy to verify that problem (\ref{prob:wsumrate-reform}) is convex with respect to any one of the optimization variables when the other optimization variables are fixed. Specifically, when the other variables are fixed, the optimal values of $\mathbf{\Sigma}_{\ell}$, $\mathbf{W}_k$, and $\mathbf{U}_k$ are given by equations (\ref{eqn:opt-sigma}), (\ref{eqn:opt-weightmatrix}), and (\ref{eqn:opt-receiver}) respectively. The optimal values of $\mathbf{\Phi}_{\ell}$ are given by $\mathbf{\Theta}_{\ell} = \mathbf{Q}_{\ell}$. When $\mathbf{\Sigma}_{\ell}$, $\mathbf{U}_k$, and $\mathbf{W}_k$ are fixed, the optimal values of $\mathbf{V}_k$ and $ \mathbf{Q}_{\ell}$ are solutions to the following optimization problem:
\begin{eqnarray}\label{sub-convex-prob}
& \displaystyle \min_{\mathbf{V}_k, \mathbf{Q}_{\ell}} &  \sum_{k=1}^K \alpha_k  \mathrm{Tr}(\mathbf{W}_k \mathbf{E}_k) + \rho\sum_{\ell=1}^L\left\|\mathbf{Q}_{\ell} -\mathbf{\Theta}_{\ell}\right\|^2_{F}\\
& \mathrm{s.t.} &  \mathrm{Tr}\left(\mathbf{\Sigma}_{\ell}^{-1}\mathbf{\Omega}_{\ell}\right) - \log\left|\mathbf{Q}_{\ell}\right| \leq C'_{\ell}  - \log\left|\mathbf{\Sigma}_{\ell}\right|, \;  \forall \ell \in \mathcal{L}, \nonumber \\
& &  \mathbf{Q}_{\ell} \succeq \mathbf{0}, \quad \forall \ell \in \mathcal{L}, \nonumber \\
& & \mathrm{Tr}\left(\mathbf{V}_k \mathbf{V}_k^{\dag}\right) \leq P_k, \quad \forall k \in \mathcal{K}, \nonumber
\end{eqnarray}
where $\mathbf{\Omega}_{\ell} = \sum_{k=1}^K\mathbf{H}_{\ell, k} \mathbf{V}_k \mathbf{V}_k^{\dag}\mathbf{H}_{\ell,k}^{\dag} + \mathbf{\Lambda}_{\ell} + \mathbf{Q}_{\ell}$.
The above problem is convex with respect to $\mathbf{V}_k$ and $ \mathbf{Q}_{\ell}$, and can be solved efficiently with
polynomial complexity. Standard convex optimization solver such as CVX~\cite{CVX2015} can be used for solving problem (\ref{sub-convex-prob}) numerically.
We summarize the proposed WMMSE-SCA algorithm for single-user compression as Algorithm \ref{alg:successive-convex}.
\begin{algorithm}[t]
\begin{algorithmic}[1]
\STATE Initialize $ \mathbf{Q}_{\ell}$ and $\mathbf{V}_k$ such that $\mathrm{Tr}\left(\mathbf{V}_k \mathbf{V}_k^{\dag}\right) = P_k$.
\STATE $\mathbf{\Sigma}_{\ell} \leftarrow \sum_{k=1}^K\mathbf{H}_{\ell, k} \mathbf{V}_k \mathbf{V}_k^{\dag}\mathbf{H}_{\ell,k}^{\dag} + \mathbf{\Lambda}_{\ell} + \mathbf{Q}_{\ell}$.
\STATE $\mathbf{U}_k \leftarrow \left(\sum_{j\neq k}\mathbf{H}_{\mathcal{L}, j} \mathbf{V}_j \mathbf{V}_j^{\dag}\mathbf{H}_{\mathcal{L},j}^{\dag} + \mathbf{\Lambda} + \mathbf{Q}\right)^{-1}\mathbf{H}_{\mathcal{L},k} \mathbf{V}_k$.
\STATE $\mathbf{W}_k\leftarrow \mathbf{I} +  \mathbf{V}^{\dag}_k\mathbf{H}^{\dag}_{\mathcal{L},k}\mathbf{U}_k$ and $\mathbf{\Theta}_{\ell} \leftarrow \mathbf{Q}_{\ell}$.
\STATE Fix $\mathbf{\Sigma}_{\ell} $, $\mathbf{U}_k$, $\mathbf{W}_k$, and $\mathbf{\Theta}_{\ell}$ solve the convex optimization problem (\ref{sub-convex-prob}). Set $(\mathbf{V}_k, \mathbf{Q}_{\ell})$ to be its optimal solution.
\STATE Repeat Steps 2--5, until convergence.
\end{algorithmic}
\caption{WMMSE-SCA Algorithm}
\label{alg:successive-convex}
\end{algorithm}

\subsection{Convergence and Complexity Analysis}

The WMMSE-SCA algorithm yields a nondecreasing sequence of objective
values for problem (\ref{prob:weight-sumrate-SU}). So the algorithm is guaranteed to converge. Moreover, it converges to
a stationary point of the optimization problem. The convergence result is stated in Theorem \ref{thm:SCAconvergence}.

\begin{thm}\label{thm:SCAconvergence}
From any initial point $\left(\mathbf{V}_k^{(0)},\mathbf{Q}^{(0)}_{\ell}\right)$, the proposed WMMSE-SCA algorithm is guaranteed to converge. The limit
point $(\mathbf{V}^*_k,\mathbf{Q}^*_{\ell})$ generated by the WMMSE-SCA algorithm
is a stationary point of the weighted sum-rate maximization problem
(\ref{prob:weight-sumrate-SU}).
\end{thm}

\begin{IEEEproof}
See Appendix \ref{sec:proof-SCAconvergence}.
\end{IEEEproof}

We point out here that Theorem \ref{thm:SCAconvergence} can also be proved following a similar procedure as that for demonstrating the convergence of WMMSE algorithm~\cite{shi2011iteratively}. Specifically, it follows from the general optimization
theory~\cite[Theorem 2.7.1]{bertsekas1999nonlinear}  that the WMMSE-SCA algorithm, which does block
coordinate descent on the reformulated problem (\ref{prob:wsumrate-reform}), converges to a stationary
point of (\ref{prob:wsumrate-reform}). Then one can show every stationary point of (\ref{prob:wsumrate-reform}) is also a stationary point of the original maximization problem (\ref{prob:weight-sumrate-SU}), thereby
establishing the claim in Theorem \ref{thm:SCAconvergence}. However such a proof is not as simple as the proof presented in this paper which utilizes the convergence result of the successive convex approximation algorithm~\cite{Scutari2014distributed}. We also emphasize the importance of the regularization term involving sum of Frobenius norms in the objective function of (\ref{prob:wsumrate-reform}). The regularization term makes the objective function in (\ref{prob:wsumrate-reform}) a strongly convex function with respect to $(\mathbf{V}_k, \mathbf{Q}_{\ell})$, therefore guaranteeing the convergence of Algorithm \ref{alg:successive-convex}.

Assuming a typical network with $K>L>N>M$, the computational complexity of the proposed WMMSE-SCA algorithm is dominated by the joint optimization of $(\mathbf{V}_k, \mathbf{Q}_{\ell})$, i.e. Step 5 of Algorithm \ref{alg:successive-convex}. Step 5 solves a convex optimization problem, which can be efficiently implemented by primal-dual interior point method with approximate complexity of $\mathcal{O}\left((KMd+LN)^{3.5}\right)$~\cite{potra2000interior}. Suppose that Algorithm \ref{alg:successive-convex} takes $T$ total number of iterations to converge, the overall computational complexity of Algorithm \ref{alg:successive-convex} is therefore $\mathcal{O}\left((KMd+LN)^{3.5}T\right)$.

\section{Joint Beamforming and Compression Optimization under Wyner-Ziv coding}
\label{sec:joint-design-WZ}
In single-user compression, the compression and decompression across different BSs take place independently. This separate processing neglects the key fact that the received signals 
$\mathbf{y}_{\ell}$  in (\ref{eqn:received-sig-Y}) are statistically correlated across the BS index $\ell$, since they are noisy observations of the same transmitted signals $\mathbf{x}_k$.
Based on this fact, Wyner-Ziv coding, which jointly decompresses the signals at the CP, is expected to be superior to the pre-link single-user compression in utilizing the limited fronthaul capacities. With fixed transmitters, the advantages of Wyner-Ziv coding have been demonstrated in~\cite{park2014fronthaul,zhou2014JSAC}. We take one step further in this section to study the problem of jointly optimizing transmit beamforming vectors and Wyner-Ziv quantization noise covariance matrices for the VMAC scheme in uplink C-RAN.

In the implementation of Wyner-Ziv coding, we decompress the quantization codeword $\hat{\mathbf{y}}_{\ell}$ sequentially from one BS to the other. To this end, we need to determine a decompression order on the BS indices $\{1,2,\ldots, L\}$. The decompression order generally affects the achievable performance of the VMAC scheme and should be optimized. However, in order to determine the optimal order that results in the largest weighted sum rate (or the maximum network utility) for the uplink C-RAN model shown in Fig.~\ref{fig:C-RAN}, we need to exhaustively search over $L!$ different decompression orders, which is impractical for large $L$. To tackle this problem, we propose a heuristic order of decompressing first the signals from the BS with larger value of
\begin{equation}
C_{\ell} - \log\left|\mathbf{H}_{\ell, \mathcal{K}} \tilde{\mathbf{K}}\mathbf{H}_{\ell,\mathcal{K}}^{\dag} + \mathbf{\Lambda}_{\ell}\right|, \quad \forall \ell \in \mathcal{L},
\end{equation}
where $\tilde{\mathbf{K}} = \mathrm{diag}\left( \{P_k \mathbf{I}\}_{k=1}^K\right)$ represents the transmit signal covariance matrix with all the users emitting independent signals across the antennas at their maximum powers. The rationale of this approach is to let signals from the BSs with either larger fronthaul capacity or lower received signal power be recovered first, then the recovered signals can sever as side information in helping the decompression of signals from other BSs. This decompression order attempts to make the quantization noise levels across the BSs small.
It is shown by simulation in the later section that the proposed heuristic approach works rather well for implementing Wyner-Ziv coding in practical uplink C-RAN when the fronthaul capacities or the received signal powers at the BSs are different.

Assume that $\pi$ is the decompression order of $\hat{\mathbf{y}}_{\ell}$ given by the heuristic approach. Denote the index set by $\mathcal{T}_{\ell} = \{\pi(1), \ldots, \pi(\ell)\}$, where $\pi(\ell)$ represents the $\ell$th component in $\pi$. Let $\mathbf{Q}_{\mathcal{T}_\ell} = \mathrm{diag}\left(\{\mathbf{Q}_{\ell}\}_{\ell\in \mathcal{T}_\ell}\right)$. The weighted sum rate maximization problem under Wyner-Ziv coding can be formulated as follows:
\begin{eqnarray}\label{prob:weight-sumrate-WZ}
\hspace{-5mm}&\displaystyle \max_{\mathbf{V}_k,\mathbf{Q}_{\ell}} &
\sum_{k=1}^K \alpha_k \log\left|\mathbf{I} + \mathbf{V}_k^{\dag} \mathbf{H}^{\dag}_{\mathcal{L},k}\mathbf{J}_k^{-1}\mathbf{H}_{\mathcal{L},k} \mathbf{V}_k \right| \\
\hspace{-5mm} &\mathrm{s.t.} &  
\displaystyle \log\frac{\left| \mathbf{\Upsilon}_{\mathcal{T}_{\ell}}  +
\mathbf{Q}_{\mathcal{T}_{\ell}}\right|}{\left| \mathbf{\Upsilon}_{\mathcal{T}_{\ell-1}}+ \mathbf{Q}_{\mathcal{T}_{\ell-1}} \right|} - \log|\mathbf{Q}_{\pi(\ell)}| \leq C_{\pi(\ell)}, \; \forall \ell \in \mathcal{L}, \nonumber \\
\hspace{-5mm} && \mathbf{Q}_{\ell} \succeq \mathbf{0}, \quad \forall \ell \in \mathcal{L}, \nonumber \\
\hspace{-5mm}&& \mathrm{Tr}\left(\mathbf{V}_k \mathbf{V}_k^{\dag}\right) \leq P_k, \quad \forall k \in \mathcal{K}, \nonumber 
\end{eqnarray}
where $ \mathbf{\Upsilon}_{\mathcal{T}_{\ell}}= \sum_{k=1}^K\mathbf{H}_{\mathcal{T}_{\ell},k} \mathbf{V}_k \mathbf{V}_k^{\dag} \mathbf{H}^{\dag}_{\mathcal{T}_{\ell},k} + \mathrm{diag}\left(\{\mathbf{\Lambda}_{\ell}\}_{\ell\in \mathcal{T}_\ell}\right)$, $\alpha_k$'s are the weights associated with the users, and $\mathbf{J}_k$ is given by either equation (\ref{eqn:J}) for the linear MMSE receiver or equation (\ref{eqn:J-SIC}) for the SIC receiver.

The above problem is again non-convex, which makes finding its global optimum challenging. To efficiently solve problem (\ref{prob:weight-sumrate-WZ}), we again utilize the successive convex approximation approach proposed in the WMMSE-SCA algorithm.  An obstacle to applying the convex approximation procedure directly to problem (\ref{prob:weight-sumrate-WZ}) lies in the Wyner-Ziv fronthaul constraint, which contains three log-determinant functions. To facilitate the utilization of the WMMSE-SCA algorithm, we reformulate problem (\ref{prob:weight-sumrate-WZ}) as an equivalent problem as follows,
\begin{eqnarray}\label{prob:wsumrate-WZ-reform1}
\hspace{-5mm}&\displaystyle \max_{\mathbf{V}_k,\mathbf{Q}_{\ell}} &
\sum_{k=1}^K \alpha_k \log\left|\mathbf{I} + \mathbf{V}_k^{\dag} \mathbf{H}^{\dag}_{\mathcal{L},k}\mathbf{J}_k^{-1}\mathbf{H}_{\mathcal{L},k} \mathbf{V}_k \right| \\
\hspace{-5mm} &\mathrm{s.t.} &  \log\left| \mathbf{\Upsilon}_{\mathcal{T}_{\ell}}  + \mathbf{Q}_{\mathcal{T}_{\ell}}\right| - \sum_{\ell \in \mathcal{T}_{\ell}}\log|\mathbf{Q}_{\ell}| \leq \sum_{\ell \in \mathcal{T}_{\ell}}C_{\ell}, \; \forall \ell \in \mathcal{L},  \nonumber \\
\hspace{-5mm} && \mathbf{Q}_{\ell} \succeq \mathbf{0}, \quad \forall \ell \in \mathcal{L}, \nonumber \\
\hspace{-5mm}&& \mathrm{Tr}\left(\mathbf{V}_k \mathbf{V}_k^{\dag}\right) \leq P_k, \quad \forall k \in \mathcal{K}, \nonumber
\end{eqnarray}
The advantage of reformulation (\ref{prob:wsumrate-WZ-reform1}) is that it has similar format as (\ref{prob:weight-sumrate-SU}), so the successive convex approximation procedure can again be used directly. Similar to the single-user case, by approximating the objective function and the fronthaul constraints in (\ref{prob:wsumrate-WZ-reform1}) with (\ref{eqn:logdeter-lowerbound}) and (\ref{eqn:rate-WMMSE}) respectively, problem (\ref{prob:wsumrate-WZ-reform1}) can be rewritten as
\begin{eqnarray}\label{prob:wsumrate-WZ-reform2}
& \displaystyle \max_{\substack{\mathbf{V}_k, \mathbf{Q}_{\ell}, \mathbf{U}_k, \\ \mathbf{W}_k, \mathbf{\Theta}_{\ell}, \mathbf{\Sigma}_{\mathcal{T}_{\ell}}}} &  \sum_{k=1}^K \alpha_k \left( \log |\mathbf{W}_k| - \mathrm{Tr}(\mathbf{W}_k \mathbf{E}_k) \right)\nonumber \\
& & \hspace{2cm} - \rho\sum_{\ell=1}^L\left\| \mathbf{Q}_{\ell} - \mathbf{\Theta}_{\ell}\right\|^2_{F} \\
& \mathrm{s.t.} & \log\left|\mathbf{\Sigma}_{\mathcal{T}_{\ell}}\right| + \mathrm{Tr}\left(\mathbf{\Sigma}_{\mathcal{T}_{\ell}}^{-1}\mathbf{\Omega}_{\mathcal{T}_{\ell}}\right) - \log|\mathbf{Q}_{\mathcal{T}_{\ell}}| \leq C'_{\mathcal{T}_{\ell}},   \nonumber\\
& & \mathbf{Q}_{\ell} \succeq \mathbf{0}, \enspace \mathbf{\Theta}_{\ell} \succeq \mathbf{0}, \quad \forall \ell \in \mathcal{L}, \nonumber \\
& & \mathrm{Tr}\left(\mathbf{V}_k \mathbf{V}_k^{\dag}\right) \leq P_k, \quad \forall k \in \mathcal{K},  \nonumber
\end{eqnarray}
where $\rho>0$ is a constant, $C'_{\mathcal{T}_{\ell}} = \sum_{\ell \in \mathcal{T}_{\ell}}\left(C_{\ell} + N\right) $, and $\mathbf{\Omega}_{\mathcal{T}_{\ell}} = \sum_{k=1}^K\mathbf{H}_{\mathcal{T}_{\ell}, k} \mathbf{V}_k \mathbf{V}_k^{\dag}\mathbf{H}_{\mathcal{T}_{\ell},k}^{\dag} + \mathrm{diag}\left(\{\mathbf{\Lambda}_{\ell} + \mathbf{Q}_{\ell}\}_{\ell\in \mathcal{T}_\ell}\right)$. Clearly, the proposed WMMSE-SCA algorithm can be applied for solving the above optimization problem. We summarize the beamforming and fronthaul compression scheme for Wyner-Ziv coding as Algorithm \ref{alg:successive-convex-WZ}.
\begin{algorithm}[t]
\begin{algorithmic}[1]
\STATE Determine a decompression order $\pi$ of $\hat{\mathbf{y}}_{\ell}$'s according to $C_{\ell} - \log\left|\mathbf{H}_{\ell, \mathcal{K}} \tilde{\mathbf{K}}\mathbf{H}_{\ell,\mathcal{K}}^{\dag} + \mathbf{\Lambda}_{\ell}\right|$.
\STATE 
Solve the optimization problem (\ref{prob:wsumrate-WZ-reform2}) using Algorithm \ref{alg:successive-convex}. Set $(\mathbf{V}_k,\mathbf{Q}_{\ell})$ to be its optimal solution.
\end{algorithmic}
\caption{Beamforming and Fronthaul Compression Optimization under Wyner-Ziv coding}
\label{alg:successive-convex-WZ}
\end{algorithm}

\section{Separate Design of Beamforming and Compression}
\label{sec:separate-design}
Although locally optimal transmit beamformers and quantization noise covariance matrices can be
found using the WMMSE-SCA algorithm for any fixed user schedule,
user priority, and channel condition, the implementation of WMMSE-SCA in
practice can be computationally intensive, especially when the channels are under fast fading or when the scheduled users in the time-frequency slots change frequently.
In this section, we aim at deriving near optimal transmit beamformers and quantization noise covariance matrices in the high
SQNR regime. The main result of this section is that a simple separate design which involves optimizing transmit beamformers for the Gaussian vector multiple-access channel at the user side and using quantizers with uniform quantization noise levels across the antennas at each BS is approximately optimal under appropriate conditions. This leads to an efficient transmit beamforming and fronthaul compression design for practical uplink C-RAN systems.

\subsection{Quantization Noise Design Under High SQNR}

The proposed approximation scheme is derived by considering the sum rate
maximization problem assuming single-user compression and assuming that
SIC is implemented at the central receiver.
Denote the transmit signal covariance matrix for the $j$th user as
$\mathbf{K}_{j} = \mathbf{V}_j\mathbf{V}^{\dag}_j$, and let
$\mathbf{K}_{\mathcal{K}} = \mathrm{diag}\left(\{\mathbf{K}_{j}\}_{j=1}^K\right)$.
The sum rate maximization problem can be formulated as
follows, 
\begin{eqnarray}\label{prob:wsumrate-SQNR}
\hspace{-5mm}& \displaystyle \max_{\mathbf{K}_{j},\mathbf{Q}_{\ell}} &
\log \frac{\left|\mathbf{H}_{\mathcal{L}, \mathcal{K}} \mathbf{K}_{\mathcal{K}} \mathbf{H}_{\mathcal{L}, \mathcal{K}}^{\dag} + \mathbf{\Lambda} + \mathbf{Q}\right|}{\left|\mathbf{\Lambda} + \mathbf{Q}\right|}  \\
\hspace{-5mm}& \mathrm{s.t.} & \log\frac{\left|\sum_{j=1}^K\mathbf{H}_{\ell, j} \mathbf{K}_{j} \mathbf{H}_{\ell, j}^{\dag} + \mathbf{\Lambda}_{\ell} + \mathbf{Q}_{\ell}\right|}{\left|\mathbf{Q}_{\ell}\right|} \leq C_{\ell}, \; \forall \ell\in \mathcal{L},\nonumber \\
\hspace{-5mm}& & \mathbf{Q}_{\ell}\succeq \mathbf{0}, \quad \forall \ell \in \mathcal{L}, \nonumber \\
\hspace{-5mm}& & \mathrm{Tr}\left(\mathbf{K}_{j} \right) \leq P_j, \quad \forall j \in \mathcal{K}, \nonumber
\end{eqnarray}
where $\mathbf{\Lambda} = \mathrm{diag}\left(\{\mathbf{\Lambda}_{\ell}\}_{\ell=1}^L\right)$ and $\mathbf{Q} = \mathrm{diag}\left(\{\mathbf{Q}_{\ell}\}_{\ell=1}^L\right)$.

In the following, we provide a justification that the optimal quantization
noise levels should be set to be uniform across the antennas at each BS for
maximizing the sum rate under high SQNR. By high SQNR, we require at least
that the received signals across all the BS antennas occupy the entire space of
receive dimensions, so implicitly enough number of users need to be scheduled, e.g., when $Kd = LN$.
Further, the received signal strength needs to be much larger than
the combined quantization and background noise level.

Mathematically, the required condition can be obtained by examining the
Karush-Kuhn-Tucker (KKT) condition for the optimization problem
(\ref{prob:wsumrate-SQNR}).  Form the Lagrangian
\begin{multline}\label{eqn:lag}
L(\mathbf{K}_{j},\mathbf{Q}_{\ell}, \lambda_{\ell},\mu_j)  =
 \log \left|\mathbf{H}_{\mathcal{L}, \mathcal{K}} \mathbf{K}_{\mathcal{K}} \mathbf{H}_{\mathcal{L}, \mathcal{K}}^{\dag} + \mathbf{\Lambda} + \mathbf{Q}\right| \\
-\log\left|\mathbf{\Lambda} + \mathbf{Q}\right|  - \sum_{\ell=1}^L \lambda_{\ell} \log\left|\sum_{j=1}^K\mathbf{H}_{\ell, j} \mathbf{K}_{j} \mathbf{H}_{\ell, j}^{\dag} + \mathbf{\Lambda}_{\ell} + \mathbf{Q}_{\ell}\right|
 \\
+ \sum_{\ell=1}^L \lambda_{\ell} \log\left|\mathbf{Q}_{\ell}\right| - \sum_{j=1}^K \mu_j \mathrm{Tr}\left(\mathbf{K}_{j} \right), 
\end{multline}
where $\lambda_{\ell}$ is the Lagrangian dual variable associated with the $\ell$th fronthaul constraint, and $\mu_j$ is Lagrangian multiplier for the $j$th transmit power constraint.

Setting $\partial L/\partial \mathbf{Q}_{\ell}$ to zero, we obtain the optimality condition as follows,
\begin{multline}\label{eqn:opt-cond-q}
\mathbf{F}_{\ell}\left(\mathbf{H}_{\mathcal{L}, \mathcal{K}} \mathbf{K}_{\mathcal{K}} \mathbf{H}_{\mathcal{L}, \mathcal{K}}^{\dag} + \mathbf{\Lambda} + \mathbf{Q}\right)^{-1}\mathbf{F}_{\ell}^T
-\left(\mathbf{\Lambda}_{\ell} + \mathbf{Q}_{\ell}\right)^{-1} \\
- \lambda_{\ell} \left(\sum_{j=1}^K\mathbf{H}_{\ell, j} \mathbf{K}_{j} \mathbf{H}_{\ell, j}^{\dag} + \mathbf{\Lambda}_{\ell} + \mathbf{Q}_{\ell}\right)^{-1} + \lambda_{\ell}  \mathbf{Q}^{-1}_{\ell} = \mathbf{0},
\end{multline}
where $\mathbf{F}_{\ell} = [\mathbf{0},\ldots,\mathbf{0},\mathbf{I}_{N},\mathbf{0},\ldots,\mathbf{0}]$ with only the $\ell$th $N\times N$ block being nonzero.
Assuming that $Kd = LN$, the received signal covariance matrix
$\mathbf{H}_{\mathcal{L}, \mathcal{K}} \mathbf{K}_{\mathcal{K}} \mathbf{H}_{\mathcal{L}, \mathcal{K}}^{\dag}$ is full rank.
Furthermore, if the overall system is to operate at reasonably high spectral
efficiency, the received signal-to-noise ratios (SNRs) are likely to be high
and the fronthaul capacities are likely to be large. In this case, we must have
$\mathbf{H}_{\mathcal{L}, \mathcal{K}} \mathbf{K}_{\mathcal{K}}
\mathbf{H}_{\mathcal{L}, \mathcal{K}}^{\dag} + \mathbf{\Lambda} + \mathbf{Q}
\gg \mathbf{\Lambda} + \mathbf{Q}$
and $\sum_{j=1}^K\mathbf{H}_{\ell, j} \mathbf{K}_{j} \mathbf{H}_{\ell, j}^{\dag} + \mathbf{\Lambda}_{\ell} + \mathbf{Q}_{\ell} \gg \mathbf{\Lambda}_{\ell}+\mathbf{Q}_{\ell}$.
Under this high SQNR condition, we argue that
$\mathbf{F}_{\ell}\left(\mathbf{H}_{\mathcal{L}, \mathcal{K}}
\mathbf{K}_{\mathcal{K}} \mathbf{H}_{\mathcal{L}, \mathcal{K}}^{\dag}
+ \mathbf{\Lambda} + \mathbf{Q}\right)^{-1}\mathbf{F}_{\ell}^T
\ll \left(\mathbf{\Lambda}_{\ell} + \mathbf{Q}_{\ell}\right)^{-1}$
and
$\left(\sum_{j=1}^K\mathbf{H}_{\ell, j} \mathbf{K}_{j} \mathbf{H}_{\ell,
j}^{\dag} + \mathbf{\Lambda}_{\ell} + \mathbf{Q}_{\ell}\right)^{-1} \ll
\mathbf{Q}^{-1}_{\ell}$, so that
the optimality condition becomes
\begin{equation}\label{eqn:opt-q}
\mathbf{Q}_{\ell} \approx \frac{\lambda_{\ell}}{1-\lambda_{\ell}}\mathbf{\Lambda}_{\ell}
\end{equation}
where $\lambda_{\ell} \in [0,1)$ is chosen to satisfy the fronthaul capacity
constraints for single-user compression. 
Following the same analysis, similar conclusion can also be obtained for the
sum rate maximization under Wyner-Ziv coding.

The above result implies that per-antenna quantizers with uniform quantization noise levels across the antennas at each BS are nearly optimal at high SQNR, although the quantization noise level may differ from BS to BS depending on the background noise levels and the fronthaul constraints. Note that this line of reasoning is very similar to the corresponding condition for the SISO case derived in \cite{zhou2014JSAC}.

It is worth emphasizing that in order to satisfy the high SQNR condition,
the number of user data streams scheduled in the system should be at least as large as
the number of receive spatial dimensions, and all these data streams must
transmit at high rate. In scenarios where the number of data
streams is less (i.e., some spatial dimensions are used for diversity instead
of multiplexing), receive beamforming at the BSs prior to quantization may
be beneficial. For example, MMSE beamforming or maximum ratio-combining may
be applied at each BS in order to reduce the number of
received dimensions before quantization. 

\subsection{Beamforming Design Under High SQNR}
We next consider the optimal transmit beamforming and power allocation
under high SQNR. Intuitively speaking, for maximizing the sum rate,
each user should align its signaling direction with the strongest
eigenmode of the effective channel and allocate power along this
direction in a ``water-filling" fashion. For this, we need to whiten
the combined quantization and background noise and interference, then
diagonalize the resulting channel to find its eigenmodes, and
iteratively perform the water-filling process among the
users~\cite{yu2004iterative}. As seen from (\ref{eqn:opt-q}), at high SQNR, the optimal quantization noise covariance matrices are  proportional to the background noise covariance matrices. Further, if $d = NL/K$, i.e., if the total number of user data streams is equal to the number of degrees of freedom in the system, then multiuser interference would be reasonably contained.

Based on the above intuition, we propose a simple beamforming design in which each user selects its transmit beamformers by ignoring the affect of fronthaul capacity limitation. Specifically, we consider the following weighted sum rate maximization problem for a Gaussian vector multiple-access channel:
\begin{eqnarray}\label{prob:wsumrate-SQNR-BF}
& \displaystyle \max_{\mathbf{K}_{j}} &   \sum_{k=1}^K \alpha_k\log \frac{\left|\sum_{j=k}^K\mathbf{H}_{\mathcal{L}, j} \mathbf{K}_{j} \mathbf{H}_{\mathcal{L}, j}^{\dag} + \mathbf{\Lambda} \right|}{\left|\sum_{j>k}^K \mathbf{H}_{\mathcal{L}, j} \mathbf{K}_{j} \mathbf{H}_{\mathcal{L}, j}^{\dag} + \mathbf{\Lambda} \right|}  \\
& \mathrm{s.t.} &  \mathrm{Tr}\left(\mathbf{K}_{j} \right) \leq P_j, \quad \forall j \in \mathcal{K}, \nonumber \\
& & \mathbf{K}_{j} \succeq \mathbf{0}, \quad  \forall j \in \mathcal{K}, \nonumber \\
& & \mathrm{rank}\left(\mathbf{K}_{j} \right) = d, \quad  \forall j \in \mathcal{K}. \nonumber
\end{eqnarray}
If the rank constraints on transmit signal covariance matrices $\mathbf{K}_{j}$ are ignored, the above problem becomes a convex optimization, and its optimum solution $\mathbf{K}^*_{j}$ can
be efficiently found through the interior-point method~\cite{vandenberghe1998determinant}. With the rank constraints applied, the problem is no longer convex, and we propose to find a set of approximately optimal transmit beamformers for user $j$ as follows. First, solve (\ref{prob:wsumrate-SQNR-BF}) with the rank constraints removed; let the optimal solution be $K_j^*$. Let $\gamma^i_{j}$ represent the $i$th largest eigenvalue of
$\mathbf{K}^*_{j}$ and $\mathbf{\Psi}^i_j$ represent its normalized eigenvector.
Then an approximately optimal transmit beamforming matrix $\mathbf{V}^*_j$ for user $j$ is given by  
\begin{equation}
\mathbf{V}^*_j = \left[\sqrt{\frac{P_j\gamma^1_j}{\Gamma_j}}\mathbf{\Psi}^1_j,\ldots,\sqrt{\frac{P_j\gamma^d_j}{\Gamma_j}}\mathbf{\Psi}^d_j\right]
\end{equation}
where $\Gamma_j = \sum_{i=1}^d \gamma^i_{j}$ represents the sum of $d$ largest of eigenvalues $\mathbf{K}^*_{j}$.

When linear MMSE receiver is employed, simply ignoring the rank constraints in the weighted sum rate maximization problem does not make it a convex optimization. In this case, one can rewrite $\mathbf{K}_j = \mathbf{V}_j\mathbf{V}^{\dag}_j$ and use the WMMSE method~\cite{christensen2008weighted, shi2011iteratively} to find the optimal beamforming vector $\mathbf{V}^*_j$.


\subsection{Separate Beamforming and Compression Design}
The above beamforming strategy together with per-antenna scalar quantizer provide us a low-complexity separate design for transmit beamforming and fronthaul compression.
With single-user compression, define
\begin{equation}
\label{eqn:C-SU}
\displaystyle C^{\mathrm{SU}}\left(\beta_{\ell}\right) = \log\frac{\left|\sum\limits_{k=1}^K\mathbf{H}_{\ell,k} \mathbf{V}_k \mathbf{V}_k^{\dag} \mathbf{H}^{\dag}_{\ell,k} + (1+ \beta_{\ell})\mathbf{\Lambda}_{\ell} \right|}{\left| \beta_{\ell}\mathbf{\Lambda}_{\ell}\right|}.
\end{equation}
To fully utilize the fronthaul capacities, the bisection search is employed to find the optimal $\beta_{\ell}$ such that $C^{\mathrm{SU}}\left(\beta_{\ell}\right) = C_{\ell}$ for $\ell=1,\ldots, L$.

With Wyner-Ziv coding, assuming without loss of generality a decoding order of $\hat{\mathbf{y}}_{\ell}$ from $1$ to $L$, define
\begin{multline}
\label{eqn:C-WZ}
\displaystyle C_j^{\mathrm{WZ}}(\beta_{1}, \ldots, \beta_j)= \\
\log\frac{\left|\sum\limits_{k=1}^K\mathbf{H}_{\mathcal{T}_{j},k} \mathbf{V}_k \mathbf{V}_k^{\dag} \mathbf{H}^{\dag}_{\mathcal{T}_{j},k} + \mathrm{diag}\left(\{(1+\beta_{\ell})\mathbf{\Lambda}_{\ell}\}_{\ell\in \mathcal{T}_{j}}\right) \right|}{\left|\mathrm{diag}\left(\{\beta_{\ell}\mathbf{\Lambda}_{\ell}\}_{\ell\in \mathcal{T}_{j}}\right)\right|}
\end{multline}
where
$\mathcal{T}_{j} = \{1,\ldots, j\}$. Different from the single-user case, the optimal $\beta_{\ell}$ in Wyner-Ziv coding is determined one after another in a successive fashion. Specifically, for $j=1,\ldots, L$, we use bisection search to find the optimal $\beta_j$ such that $C_j^{\mathrm{WZ}}(\beta_{1}, \ldots, \beta_{j-1}, \beta_j)  = \sum_{\ell=1}^{j} C_{\ell}$ assuming that the values of $\beta_{1}, \ldots, \beta_{j-1}$ are fixed and as determined by the previous $j-1$ bisection searches.

\begin{algorithm}[t]
\begin{algorithmic}[1]
\STATE Solve problem (\ref{prob:wsumrate-SQNR-BF}) with the
rank constraints removed and set $\mathbf{K}^*_{j}$ to be its optimal solution.
\STATE Perform eigenvalue decomposition on $\mathbf{K}^*_{j}$ to obtain its normalized eigenvalues $\gamma^i_j$ and eigenvectors $\mathbf{\Psi}^i_j$. Set $ \mathbf{V}^*_j = \left[\sqrt{\frac{P_j\gamma^1_j}{\Gamma_j}}\mathbf{\Psi}^1_j,\ldots,\sqrt{\frac{P_j\gamma^d_j}{\Gamma_j}}\mathbf{\Psi}^d_j\right]$. 
\STATE Under single-user compression, use bisection method in $[\beta^{\min}_{\ell}, \beta^{\max}_{\ell}]$ to solve for $\beta_{\ell}$ in $C^{\mathrm{SU}}\left(\beta_{\ell}\right) = C_{\ell}$ independently for $\ell=1,\ldots,L$; Under Wyner-Ziv coding, use bisection in  $[\beta^{\min}_{j}, \beta^{\max}_{j}]$ to solve for $\beta_{j}$ in $C_j^{\mathrm{WZ}}(\beta_{1}, \ldots, \beta_{j-1}, \beta_j)  = \sum_{\ell=1}^{j} C_{\ell}$ for $j=1,\ldots,L$ successively with the values of $\beta_{1}, \ldots, \beta_{j-1}$ fixed and as determined by the previous $j-1$ bisection searches.
\STATE Set $\mathbf{Q}_{\ell} = \beta_{\ell}\mathbf{\Lambda}_{\ell}$ for $\ell=1,\ldots,L$.
\end{algorithmic}
\caption{Separate Beamforming and Fronthaul Compression Design}
\label{alg:approx-scheme}
\end{algorithm}

The separate transmit beamforming and fronthaul compression design scheme is summarized as Algorithm \ref{alg:approx-scheme}. In the Step 3 of Algorithm \ref{alg:approx-scheme}, the values of $\beta^{\min}_{\ell}$ and $\beta^{\max}_{\ell}$ can be found using the same way in~\cite{zhou2014JSAC}. Specifically, under single-user compression, initialize $\beta_{\ell} = 1$ for $\ell=1,\ldots, L$, and keep updating $\beta_{\ell} = 2 \beta_{\ell}$ until 
$C^{\mathrm{SU}}\left(\beta_{\ell}\right) \leq C_{\ell}$
is satisfied. Then, we set $\beta^{\min}_{\ell} = 0$ and $\beta^{\max}_{\ell} = \beta_{\ell}$ for $\ell=1,\ldots,L$. Similar procedure can also be used in the case of Wyner-Ziv coding.


There are two differences between the joint design scheme and the separate design scheme. First, in the joint design, transmit beamforming are chosen to be fronthaul-aware, while the impact of limit fronthaul is ignored in the separate design. Second, in the joint design, the quantization is performed on the received signal vector across all the receive antennas at each BS while separate design adopts per-antenna quantization on each receive antenna of the BSs. It is shown by simulation in the next section that the separate design performs very well in the high SQNR regime. In other regimes, the difference between the joint design and separate design represents a tradeoff between complexity and performance in implementing uplink C-RAN.

We remark that when the high SQNR condition is not satisfied, the optimal beamforming in uplink C-RAN should be fronthaul-aware. For example, consider a two-layer heterogenous C-RAN system with both pico BSs and macro BSs serving as relay nodes. The fronthaul capacity of the macro BS is typically much larger than that of the pico BS. Therefore, users are more likely to form their transmit beamformer pointing toward the receive antennas at the macro BSs rather than the pico BSs. Under this scenario, both of the channel strength between the users and the BSs and the fronthaul capacities between the BSs and the CP should be taken into account in the beamforming design in order to maximize the network throughput.

From the computational complexity point of view, the separate design is significantly superior to  the joint design. Algorithm \ref{alg:approx-scheme} involves solving a single convex optimization problem (\ref{prob:wsumrate-SQNR-BF}) plus a bisection search, as compared to iteratively solving a series of convex optimization problems (\ref{sub-convex-prob}) or (\ref{prob:wsumrate-WZ-reform2}) as in the WMMSE-SCA algorithm.

\section{Simulation Results}
\label{sec:simu}

\subsection{Single-Cluster Network}

\begin{table}[!t]
\caption{Multicell Network System Parameters}
\centering
\begin{tabular}{|c|c|}
\hline
 Cellular Layout & Hexagonal, $19$-cell, $3$ sectors/cell  \\ \hline
 BS-to-BS Distance & $500$ m    \\ \hline
 Frequency Reuse  & $1$     \\ \hline
 Channel Bandwidth & $10$ MHz    \\ \hline
 Number of Users per Sector  & $20$   \\ \hline
 Total Number of Users  & $420$   \\ \hline
 Max Transmit Power & $23$ dBm   \\ \hline
 Antenna Gain & 14 dBi \\ \hline
 Background Noise  & $-169$ dBm/Hz \\ \hline
 Noise Figure & $7$ dB \\ \hline
 Tx/Rx Antenna No. & $2\times 2$ \\ \hline
 Distance-dependent Path Loss & $128.1+ 37.6 \log_{10}(d)$ \\ \hline
 Log-normal Shadowing &  $8$ dB standard deviation  \\ \hline
 Shadow Fading Correlation &  $0.5$  \\ \hline
 Cluster Size & $7$ cells ($21$ sectors)  \\ \hline
 Scheduling Strategy & WMMSE based scheduling  \\ \hline
\end{tabular}
\label{table:singlecluster-parameter}
\end{table}

In this section, the performances of the proposed WMMSE-SCA schemes with different compression strategies (i.e., Wyner-Ziv coding and single-user compression) and different receiving schemes (i.e., linear MMSE receiver and SIC receiver) are evaluated on a $19$-cell $3$-sector/cell wireless network setup with center $7$ cells (i.e., $21$ sectors) forming a cooperating cluster. The users are randomly located and associated with the strongest BS. The proposed WMMSE-SCA algorithm is applied to all the users within the cluster,
which automatically schedules the users with non-zero beamforming vectors. Each BS is equipped with $N=2$ antennas, each user is equipped with $M=2$ antennas, and each user sends one data stream (i.e., $d=1$) to the CP. Perfect channel estimation is assumed, and the CSI is made available to all BSs and to the CP. Various algorithms are run on fixed set of channels. Detailed system parameters are outlined in Table \ref{table:singlecluster-parameter}.

\begin{figure}[t]
\centering
\includegraphics[width=0.46\textwidth]{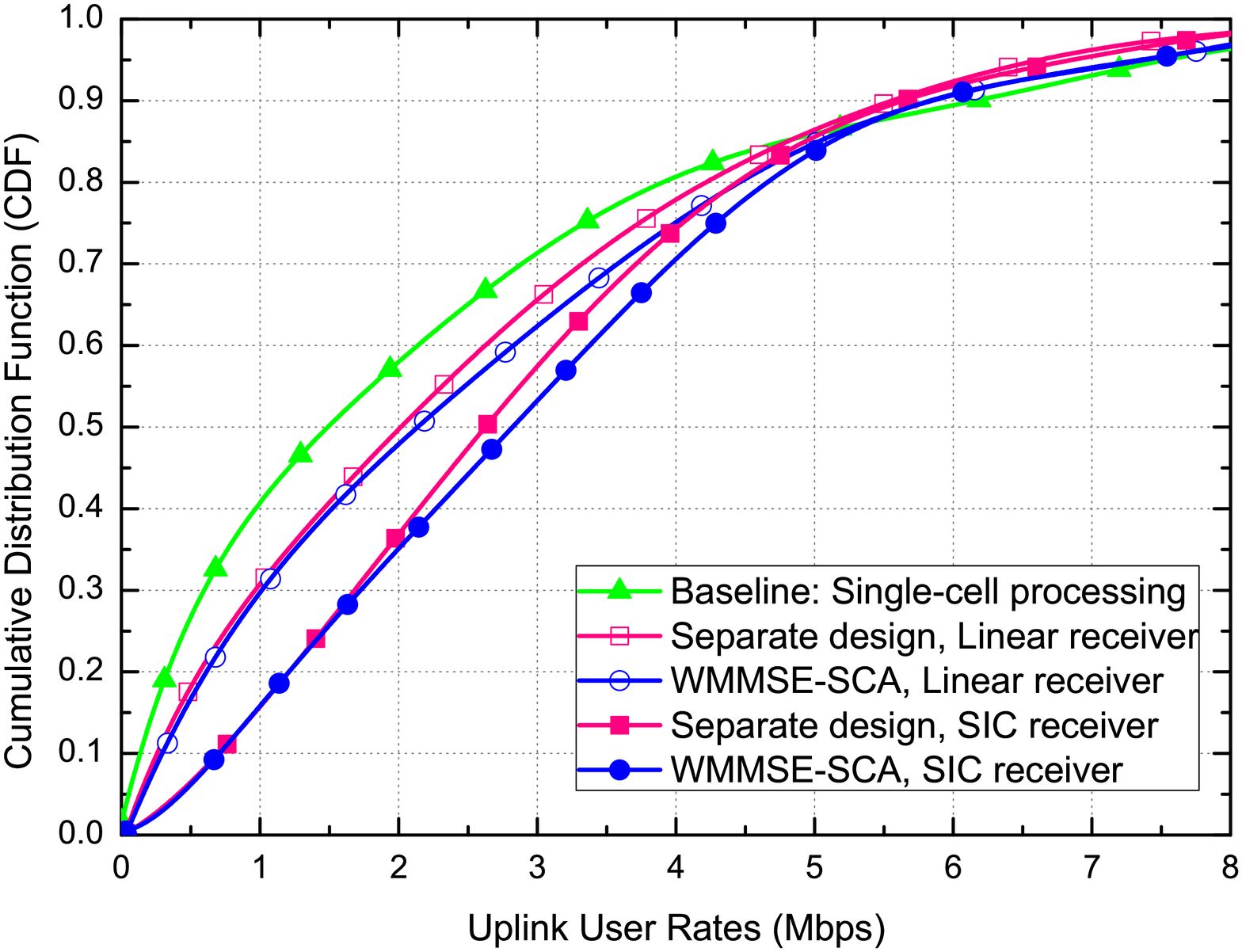}
\caption{Cumulative distribution of user rates with single-user compression for a 19-cell network with center 7 cells forming a single cluster under the fronthaul capacity of $120$Mbps per sector.}
\label{fig:cdf}
\end{figure}

\begin{figure}[t]
\centering
\includegraphics[width=0.46\textwidth]{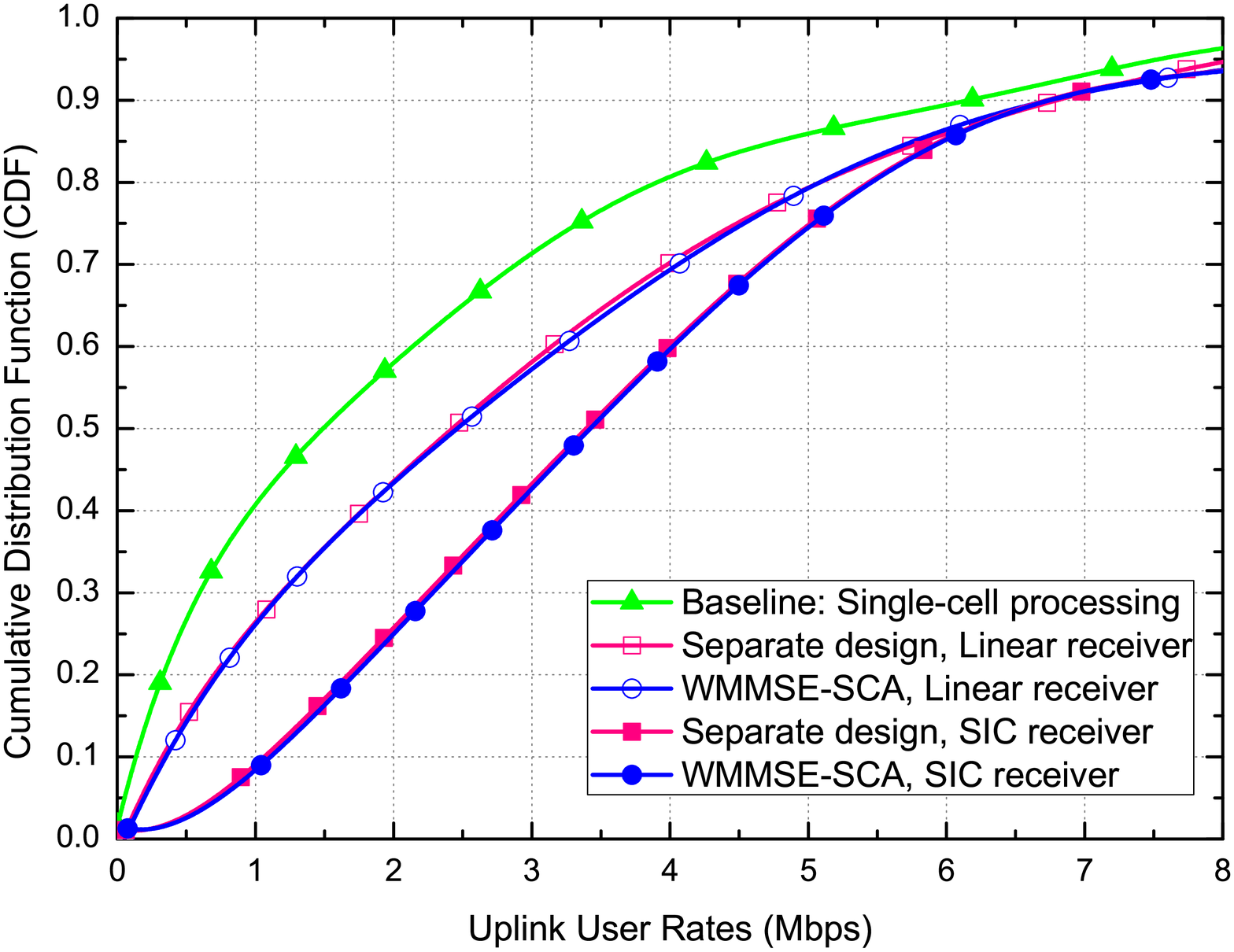}
\caption{Cumulative distribution of user rates with single-user compression for a 19-cell network with center 7 cells forming a single cluster under the fronthaul capacity of $320$Mbps per sector.}
\label{fig:cdf_SIC}
\end{figure}

Under single-user compression, Fig.~\ref{fig:cdf} and Fig.~\ref{fig:cdf_SIC} compare the performance of the WMMSE-SCA and separate design schemes
implemented either with SIC (labeled as ``SIC receiver" in the figures) or without SIC (labeled as ``linear receiver" in the figures) at the receiver under two different fronthaul constraints.
It is shown that both the WMMSE-SCA scheme and the separate design scheme significantly
outperform the baseline scheme without multicell processing.
Fig.~\ref{fig:cdf} and Fig.~\ref{fig:cdf_SIC} show that the SIC receiver achieves significant gain as compared to the linear receiver.
The performance improvement is more significant for the users with low rate. 

\begin{figure}[t]
\centering
\includegraphics[width=0.465\textwidth]{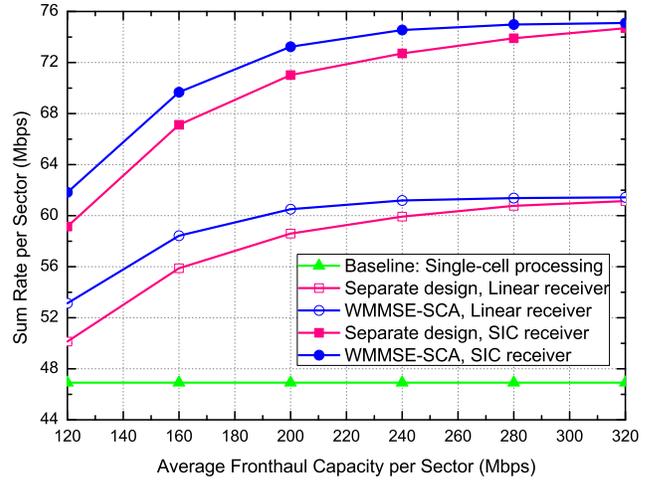}
\caption{Per-cell sum rate vs. average per-sector fronthaul capacity for single-user compression with linear receiver and with SIC receiver for a 19-cell network with center 7 cells forming a single cluster.}
\label{fig:sumrate}
\end{figure}

\begin{figure}[t]
\vspace{.3mm}
\centering
\includegraphics[width=0.46\textwidth]{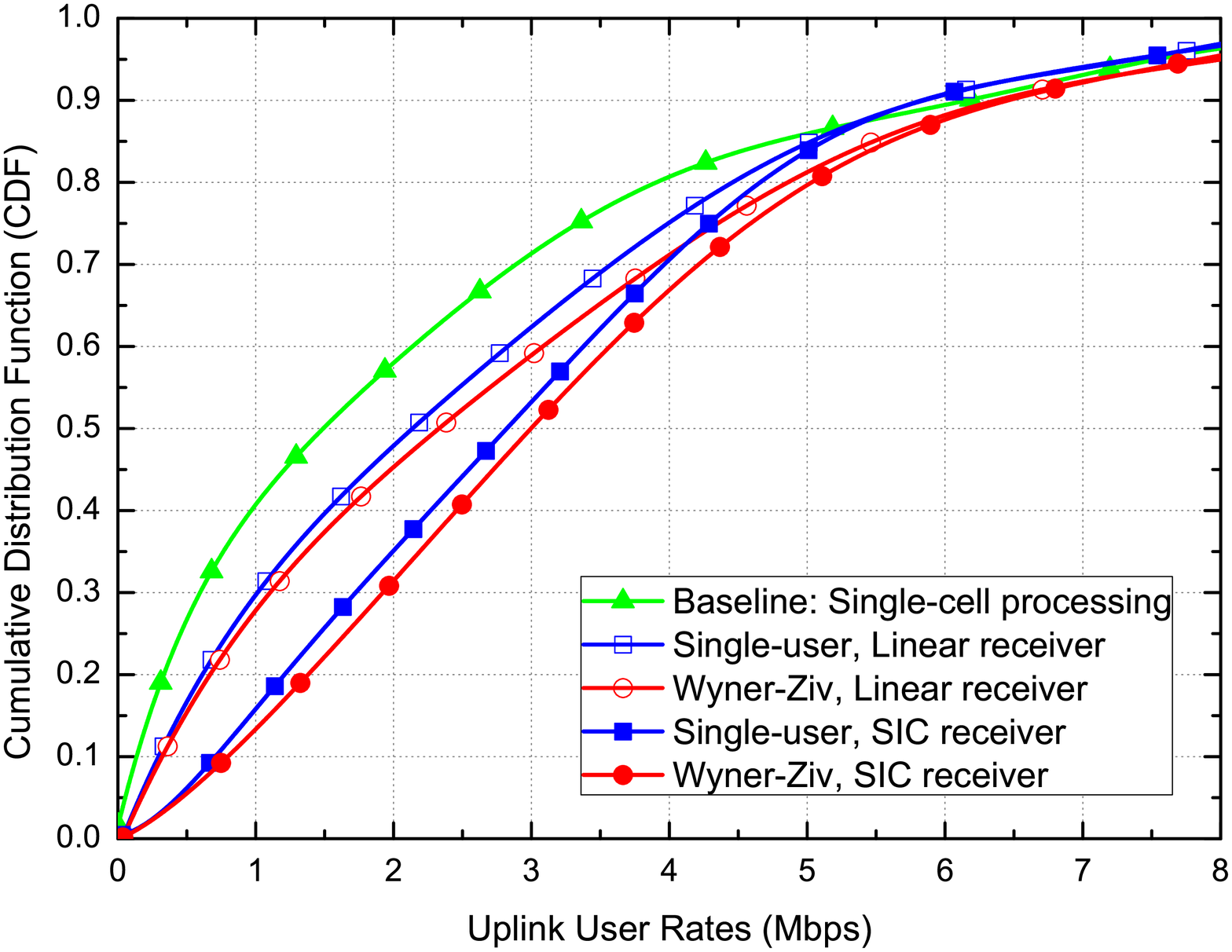}
\caption{Cumulative distribution of user rates with either single-user compression or Wyner-Ziv coding for a 19-cell network with center 7 cells forming a single cluster under the fronthaul capacity of $120$Mbps per sector.}
\label{fig:cdf_WZSU_12}
\end{figure}

\begin{figure}[t]
\centering
\includegraphics[width=0.46\textwidth]{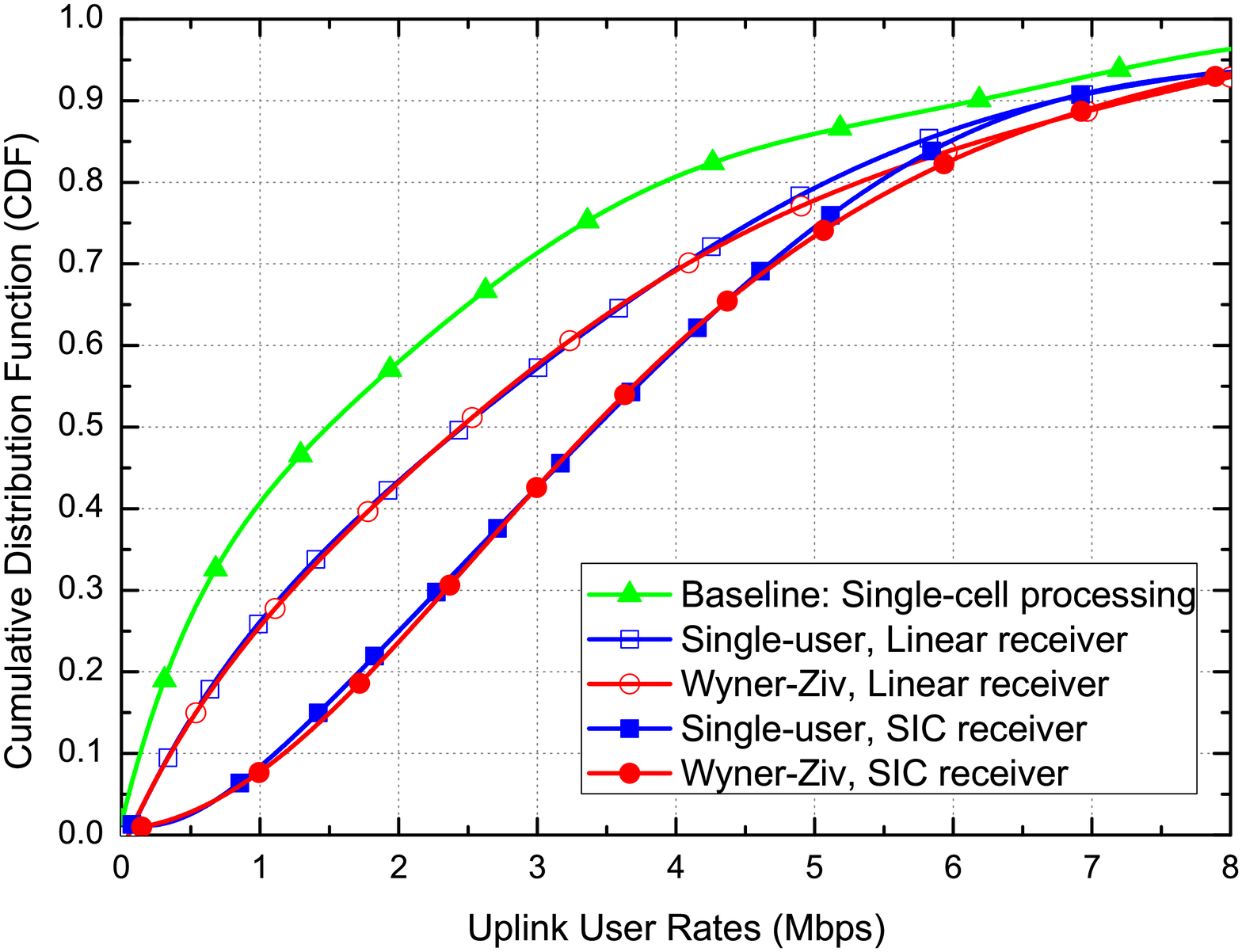}
\caption{Cumulative distribution of user rates with either single-user compression or Wyner-Ziv coding using WMMSE-SCA algorithm for a 19-cell network with center 7 cells forming a single cluster under the fronthaul capacity of $320$Mbps per sector.}
\label{fig:cdf_WZSU_32}
\end{figure}

\begin{figure}[t]
\hspace{.5mm}
\centering
\includegraphics[width=0.465\textwidth]{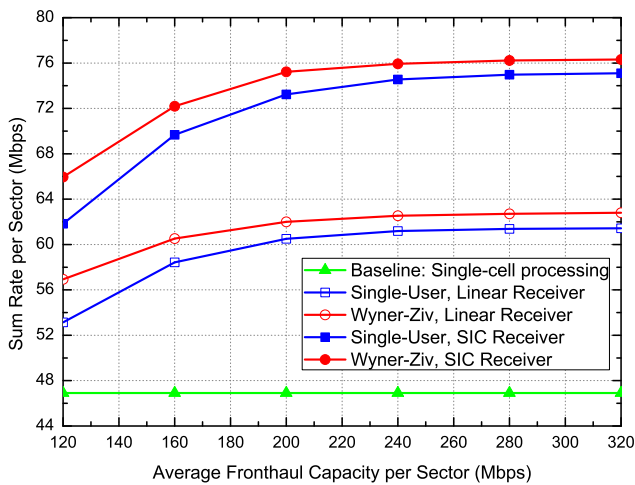}
\caption{Per-cell sum rate vs. average per-cell fronthaul capacity with either single-user compression or Wyner-Ziv coding using WMMSE-SCA algorithm for a 19-cell network with center 7 cells forming a single cluster.}
\label{fig:sumrate_WZvsSU}
\end{figure}

To further compare the performance of the proposed two schemes, Fig.~\ref{fig:sumrate} plots the average per-cell sum rate of the WMMSE-SCA scheme and the low-complexity separate design as a function of the fronthaul capacity. As the fronthaul capacity increases, the performance
gap between these two schemes becomes smaller.
This demonstrates the approximate optimality for separate design of transmit beamforming and fronthaul compression in the high SQNR regime.

Fig.~\ref{fig:cdf_WZSU_12} and Fig.~\ref{fig:cdf_WZSU_32} show the CDF curves of user rates for the WMMSE-SCA scheme implemented with four different choices of coding schemes: with either single-user or Wyner-Ziv compression at the BSs and with either linear MMSE or SIC receiver at the CP. It can be seen from Fig.~\ref{fig:cdf_WZSU_12} that under the fronthaul capacity of $120$Mbps, single-user compression with SIC receiver significantly improves the performance of linear MMSE receiver. Further gain on performance can be obtained if one replaces single-user compression by Wyner-Ziv coding. As the capacity of fronthaul increases to $320$Mbps, as shown in Fig.~\ref{fig:cdf_WZSU_32}, the gain due to Wyner-Ziv coding becomes negligible. In this high fronthaul scenario,  SIC receiver still achieves a very large gain. 

In order to quantify the performance gain brought by Wyner-Ziv coding and SIC receiver, Fig.~\ref{fig:sumrate_WZvsSU} shows the
average per-cell sum rate obtained by different schemes as the average capacity of fronthaul increases. It is observed that, under fronthaul capacity of $320$Mpbs, SIC receiver and Wyner-Ziv coding outperform the linear receiver and single-user compression respectively. But the performance improvement of SIC receiver upon linear receiver is much larger than the gain of Wyner-Ziv coding over single-user compression.

\subsection{Multi-Cluster Network}

\begin{table}[t]
\centering
\caption{Multi-Cluster Network Parameters}
\label{table:multclus-parameter}
\begin{tabular}{|c|c|}
\hline
 Cellular Layout & Hexagonal  \\ \hline
 BS-to-BS Distance & $200$ m    \\ \hline
 Frequency Reuse  & $1$     \\ \hline
 Channel Bandwidth & $10$ MHz    \\ \hline
 Number of Users per Cell  & $10$   \\ \hline
 Number of Cells  & $65$   \\ \hline
 Total Number of Users  & $650$   \\ \hline
 Max Transmit Power & $23$ dBm   \\ \hline
 Antenna Gain & 14 dBi \\ \hline
 Background Noise  & $-169$ dBm/Hz \\ \hline
 Noise Figure & $7$ dB \\ \hline
 Tx Antenna No. & $2$ \\ \hline
 Rx Antenna No. & $4$ \\ \hline
 Distance-dependent Path Loss & $128.1+ 37.6 \log_{10}(d)$ \\ \hline
 Log-normal Shadowing &  $8$ dB standard deviation  \\ \hline
 Shadow Fading Correlation &  $0.5$  \\ \hline
 Scheduling Strategy & Round-robin  \\ \hline
\end{tabular}
\end{table}

\begin{figure}[t]
\centering
\includegraphics[width=0.46\textwidth]{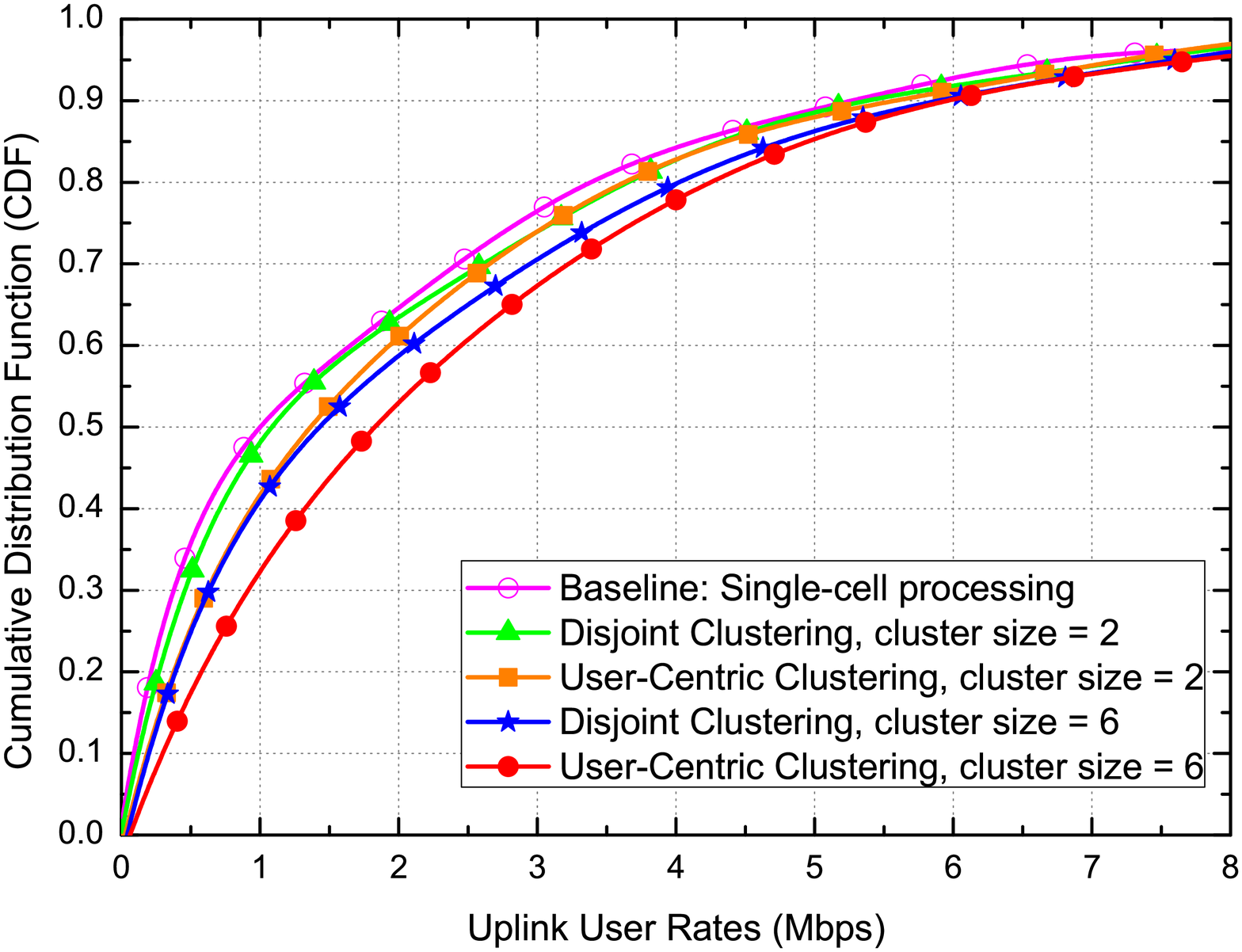}
\caption{Cumulative distribution of user rates for the WMMSE-SCA algorithm with single-user compression under the average fronthaul capacity of $120$Mbps with either disjoint or user-centric clustering for a multi-cluster network.}
\label{fig:cdf_DisUEC_12}
\end{figure}

\begin{figure}[t]
\centering
\includegraphics[width=0.46\textwidth]{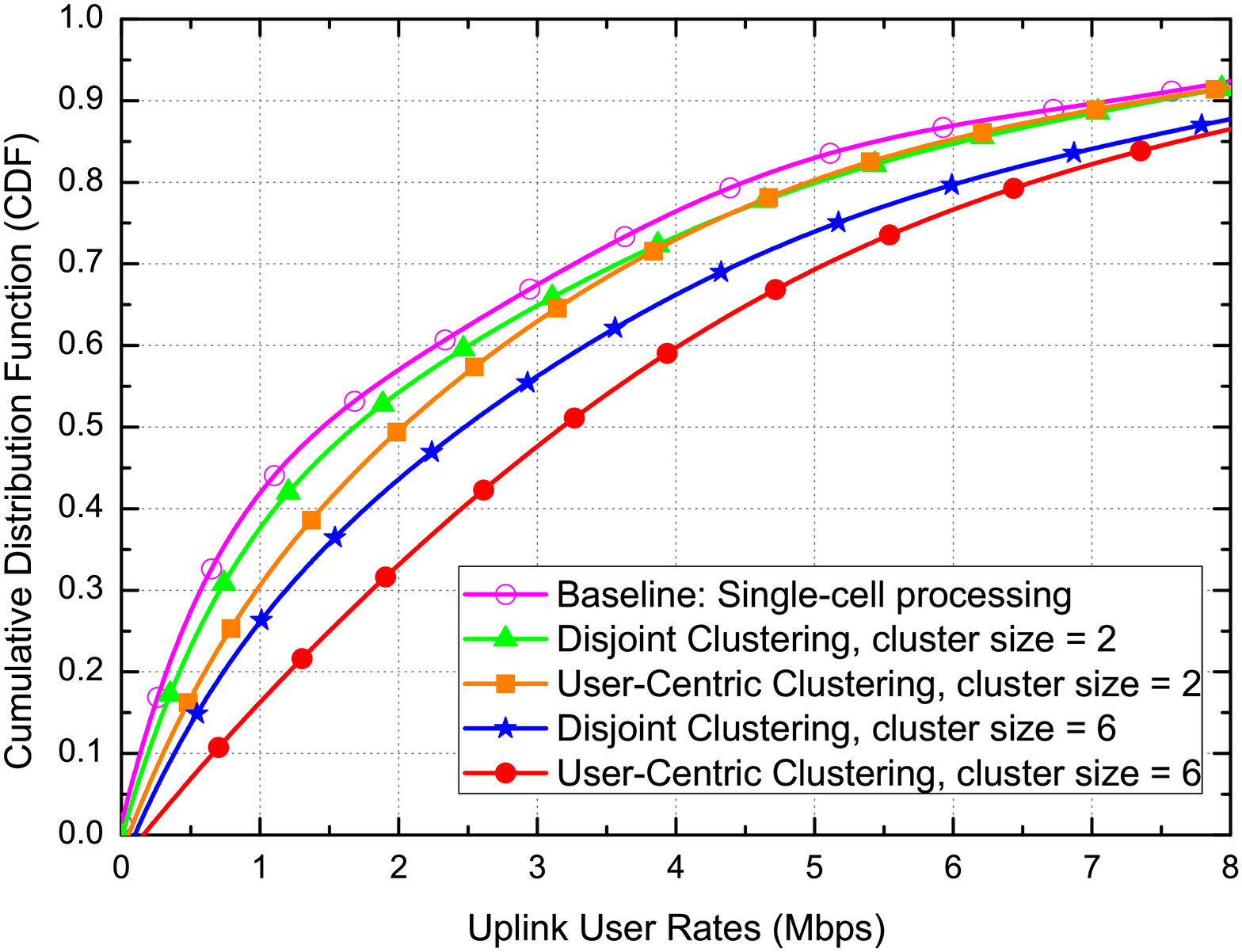}
\caption{Cumulative distribution of user rates for the WMMSE-SCA algorithm with single-user compression under the average fronthaul capacity of $360$Mbps with either disjoint or user-centric clustering for a multi-cluster network.}
\label{fig:cdf_DisUEC_32}
\end{figure}

\begin{figure}[t]
\centering
\includegraphics[width=0.465\textwidth]{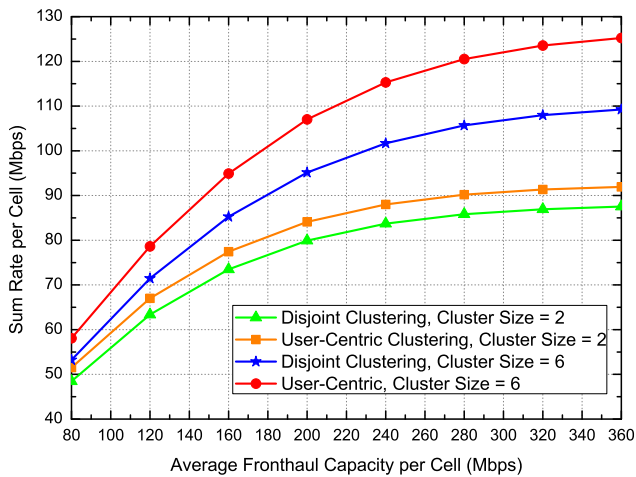}
\caption{Per-cell sum rate vs. average per-cell fronthaul capacity of the WMMSE-SCA algorithm with single-user compression for a multi-cluster network under different clustering strategies and different cluster size.}
\label{fig:sumrate_fronthaul}
\end{figure}

The performance of the proposed WMMSE-SCA scheme is further evaluated for a large-scale multicell network with $65$ cells and $10$ mobile users randomly located within each cell. The BS-to-BS distance is set to be $200$m, each user is equipped with $2$ transmit antennas, and each BS is equipped with $4$ receive antennas. The channel is assumed to be flat-fading. Round-robin user scheduling is used on a per-cell
basis and system is operated with loading factor $0.5$, i.e., in each time slot, BS schedules two users.
Detailed system parameters are outlined in Table \ref{table:multclus-parameter}.
Two different clustering strategies, i.e., disjoint clustering~\cite{3GPP11, peng2015fronthaul} and user-centric clustering~\cite{ng2010linear, dai2014sparse}, are applied to form clusters within the network. Disjoint clustering partitions the BSs in the network into nonoverlapping sets of cooperating clusters. In user-centric clustering, each user chooses a set of nearest BSs to form a cooperation cluster, and cooperating clusters overlap, which makes the implementation of Wyner-Ziv coding and SIC receiver under fronthaul capacity constraints (\ref{eqn:fronthaul-contraint-WZ}) more difficult. Therefore, for fair comparison, we only consider here the case where single-user compression and linear MMSE receiver are employed.

Fig.~\ref{fig:cdf_DisUEC_12} and Fig.~\ref{fig:cdf_DisUEC_32} show the CDF plots of user rates achieved with both disjoint clustering and user-centric clustering with WMMSE-SCA. It is clear that with optimized beamforming and fronthaul compression, the user-centric clustering significantly improves over disjoint clustering, and both of these two schemes improve as the cluster size increases. As the capacity of fronthaul links increases from $120$Mbps to $360$Mbps, the performance gap between the two clustering schemes becomes larger. Further, for disjoint clustering, increasing the cluster size from $2$ to $6$ achieves $60$\% performance improvement for the $50$-percentile rate. This gain doubles when we further replace  disjoint clustering with user-centric clustering.

Fig.~\ref{fig:sumrate_fronthaul} plots the average per-cell sum rate as the fronthaul capacity increases. The result again shows that user-centric clustering achieves significant performance gain over disjoint clustering. When cluster size increases to $6$, to achieve per-cell sum rate of $110$Mbps, disjoint clustering needs fronthaul capacity of $360$Mbps, while user-centric needs $220$Mbps, which is more than $60$\% improvement on the fronthaul requirement.

\begin{figure}[t]
\centering
\includegraphics[width=0.46\textwidth]{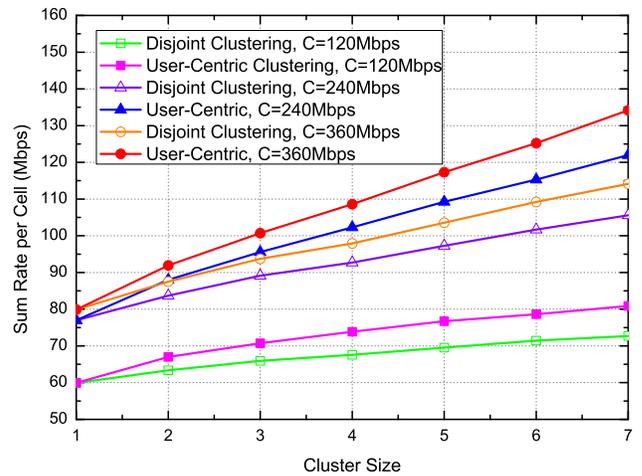}
\caption{Per-cell sum rate vs. cluster size for the WMMSE-SCA algorithm with single-user compression for a multi-cluster network under different clustering strategies and different fronthaul capacity constraints.}
\label{fig:clustersize}
\end{figure}

Finally, the performance of the two different clustering strategies are compared as a function of cluster size in Fig.~\ref{fig:clustersize}. It is shown that for both disjoint clustering and user-centric clustering, the average per-cell sum rate increases as either the cluster size or fronthaul capacity increases. As expected, user-centric clustering always outperforms disjoint clustering. If we compare the performance of disjoint clustering with fronthaul capacity of $360$Mbps with user-centric clustering with fronthaul capacity of $240$Mbps, we see that even with $120$Mbps lower fronthaul capacity, user-centric clustering already achieves higher per-cell sum rate. This improvement on per-cell sum rate becomes larger as the cluster size increases.

\section{Conclusion}
\label{sec:conclusion}

This paper studies the fronthaul compression and transmit beamforming
design for an uplink MIMO C-RAN system. From algorithm design perspective,
we propose a novel WMMSE-SCA algorithm to
efficiently optimize the transmit beamformer and quantization
noise covariance matrix jointly for maximizing the weighted sum rate with either Wyner-Ziv coding or single-user compression.
Further, we propose a separate design consisting of
transmit beamforming optimized for the Gaussian vector multiple-access channel without accounting for compression together with scalar quantization with uniform quantization noise levels across the antennas at each BS. This low-complexity separate design is shown to be near optimal for maximizing
the weighted sum rate when the SQNR is high.
The performance of optimized beamforming and fronthaul compression is evaluated for practical multicell networks with different compression strategies, different receiving schemes, and different clustering methods. Numerical results show that, with optimized beamforming and fronthaul compression, C-RAN can significantly improve the overall performance of MIMO cellular
networks. Most of the performance gain are due to the implementation of SIC at the central receiver. Finally, user-centric clustering significantly outperforms disjoint clustering in terms of fronthaul capacity saving.

\appendices

\section{Proof of Theorem \ref{thm:SCAconvergence}}
\label{sec:proof-SCAconvergence}
The proof of Theorem \ref{thm:SCAconvergence} is a direct application of the convergence result of the successive convex approximation algorithm~\cite{Scutari2014distributed}. Let $\mathbf{V} = \mathrm{diag}\left(\{\mathbf{V}_k\}_{k=1}^K\right)$. Define the objective function and fronthaul constraints in problem (\ref{prob:weight-sumrate-SU}) to be
\begin{align*}
 f(\mathbf{V}, \mathbf{Q}) = &\sum_{k=1}^K \alpha_k \log\left|\mathbf{I} + \mathbf{V}_k^{\dag} \mathbf{H}^{\dag}_{\mathcal{L},k}\mathbf{J}_k^{-1}\mathbf{H}_{\mathcal{L},k} \mathbf{V}_k \right|, \\
g_{\ell}(\mathbf{V}, \mathbf{Q})  =  &\log\frac{\left|\sum_{k=1}^K\mathbf{H}_{\ell, k} \mathbf{V}_k \mathbf{V}_k^{\dag}\mathbf{H}_{\ell,k}^{\dag} + \mathbf{\Lambda}_{\ell} + \mathbf{Q}_{\ell}\right|}{\left|\mathbf{Q}_{\ell}\right|} - C_{\ell} , 
\end{align*}
where
$\mathbf{J}_k  = \sum_{j \neq k}^K \mathbf{H}_{\mathcal{L},j} \mathbf{V}_j \mathbf{V}_j^{\dag}\mathbf{H}^{\dag}_{\mathcal{L},j} +  \mathbf{\Lambda} + \mathbf{Q}$ for the linear receiver or
$\mathbf{J}_k  = \sum_{j > k}^K \mathbf{H}_{\mathcal{L},j} \mathbf{V}_j \mathbf{V}_j^{\dag}\mathbf{H}^{\dag}_{\mathcal{L},j} +  \mathbf{\Lambda} + \mathbf{Q}$
for the SIC receiver.

At the $t$th iteration, assume that the output of WMMSE-SCA  algorithm is $(\mathbf{V}^{t}, \mathbf{Q}^{t})$. Putting $(\mathbf{V}^{t}, \mathbf{Q}^{t})$ into equations (\ref{eqn:opt-sigma}) and (\ref{eqn:opt-weightmatrix}) gives
\begin{align*}
\mathbf{\Sigma}^t_{\ell} &= \sum_{k=1}^K\mathbf{H}_{\ell, k} \mathbf{V}^t_k (\mathbf{V}^t_k)^{\dag}\mathbf{H}_{\ell,k}^{\dag} + \mathbf{\Lambda}_{\ell} + \mathbf{Q}^t_{\ell}, \\
\mathbf{W}^t_k &= \mathbf{I} + (\mathbf{V}^{t}_k)^{\dag}\mathbf{H}^{\dag}_{\mathcal{L},k} \mathbf{U}^t_k,
\end{align*}
where
\begin{equation*}
\mathbf{U}^t_k = \left(\sum_{j\neq k}\mathbf{H}_{\mathcal{L}, j} \mathbf{V}^t_j (\mathbf{V}^t_j)^{\dag}\mathbf{H}_{\mathcal{L},j}^{\dag} + \mathbf{\Lambda} + \mathbf{Q}^t\right)^{-1}\mathbf{H}_{\mathcal{L},k} \mathbf{V}^t_k.
\end{equation*}
Then the objective function and fronthaul constraints in problem (\ref{prob:wsumrate-reform}) can be written as
\begin{align*}
\tilde{f}\left(\{\mathbf{V}, \mathbf{Q}\}, \{\mathbf{V}^{t}, \mathbf{Q}^{t}\}\right) = &\sum_{k=1}^K \alpha_k \left(\log |\mathbf{W}^{t}_k|-\mathrm{Tr}\left(\mathbf{W}^{t}_k \mathbf{E}_k \right)\right) \\
& - \rho\sum_{\ell=1}^L\left\|\mathbf{Q}_{\ell} -\mathbf{Q}^t_{\ell}\right\|^2_{F}, \\
\tilde{g}_{\ell}\left(\{\mathbf{V}, \mathbf{Q}\}, \{\mathbf{V}^{t}, \mathbf{Q}^{t}\}\right)  = &
\log\left|\mathbf{\Sigma}^{t}_{\ell}\right| + \mathrm{Tr}\left((\mathbf{\Sigma}^{t}_{\ell})^{-1}\mathbf{\Omega}_{\ell}\right) \\
& -\log\left|\mathbf{Q}_{\ell}\right| - C_{\ell} -N, 
\end{align*}
where
\begin{multline*}
\mathbf{E}_k = \left(\mathbf{I} - (\mathbf{U}^t_k)^{\dag}\mathbf{H}_{\mathcal{L},k}\mathbf{V}_k\right)\left(\mathbf{I} - (\mathbf{U}^t_k)^{\dag}\mathbf{H}_{\mathcal{L},k}\mathbf{V}_k\right)^{\dag} \\
+ (\mathbf{U}^t_k)^{\dag}\left(\sum_{j\neq k}^K \mathbf{H}_{\mathcal{L},j} \mathbf{V}_j \mathbf{V}_j^{\dag}\mathbf{H}^{\dag}_{\mathcal{L},j} + \mathbf{\Lambda} + \mathbf{Q}\right)\mathbf{U}^t_k,
\end{multline*}
and $\mathbf{\Omega}_{\ell} = \sum_{k=1}^K\mathbf{H}_{\ell, k} \mathbf{V}_k \mathbf{V}_k^{\dag}\mathbf{H}_{\ell,k}^{\dag} + \mathbf{\Lambda}_{\ell} + \mathbf{Q}_{\ell}$.

We now observe that the WMMSE-SCA algorithm is actually a special case of the general successive convex approximation (SCA) method, with $\tilde{f}$ and $\tilde{g}_{\ell}$ being the convex approximation functions of $f$ and $g_{\ell}$ respectively. Furthermore, based on the fact that $\tilde{f}$ is strictly convex over $(\mathbf{V}, \mathbf{Q})$ and the result of \cite[Lemma 3.1]{Beck2010sequential}, it can be shown that $\tilde{f}$ is uniformly strongly convex over $(\mathbf{V}, \mathbf{Q})$. We point out here that the regularization term, $\rho\sum_{\ell=1}^L\left\|\mathbf{Q}_{\ell} -\mathbf{Q}^t_{\ell}\right\|^2_{F}$, plays a key role in making $\tilde{f}$ strongly convex.

Define
\begin{equation}
\label{def:feasible-region}
\mathcal{X} \triangleq \left\{\left(\mathbf{V}, \mathbf{Q}\right)\left|
\begin{array}{l}
  \mathbf{Q}_{\ell} \succeq \mathbf{0}, \enspace \forall \ell \in \mathcal{L} \\
  \mathrm{Tr}\left(\mathbf{V}_k \mathbf{V}_k^{\dag}\right) \leq P_k, \enspace \forall k \in \mathcal{K}
\end{array}\right.
\right\}
\end{equation}
and
\begin{equation}
\label{def:feasible-region-2}
\mathcal{Y} \triangleq \left\{\left(\mathbf{V}, \mathbf{Q}\right)\left|
\begin{array}{l}
  g_{\ell}\left(\mathbf{V}, \mathbf{Q}\right) \leq 0, \enspace \forall \ell \in \mathcal{L} \\
  \left(\mathbf{V}, \mathbf{Q}\right) \in \mathcal{X}
\end{array}\right.
\right\}
\end{equation}
We summarize the conditions that are satisfied for the functions $f$, $g_{\ell}$, $\tilde{f}$ and $\tilde{g}_{\ell}$ as follows:
\begin{enumerate}
  \item $\mathcal{X}$ is closed and convex (and nonempty);
  \item $f$ and $g_{\ell}$ are continuous and differentiable on $\mathcal{X}$, and $\nabla f$ is Lipschitz continuous on $\mathcal{X}$;
  \item $\tilde{f}\left(\cdot, \mathbf{y}\right)$ is uniformly strongly convex on $\mathcal{X}$ for all $\mathbf{y}\in \mathcal{Y} $ with some positive constant;
  \item $\tilde{f}\left(\cdot, \cdot\right)$ is continuous on $\mathcal{X} \times \mathcal{Y}$ and $\nabla_{\mathbf{y}} \tilde{f}\left(\mathbf{y}, \mathbf{y}\right) = \nabla_{\mathbf{y}} f\left(\mathbf{y}\right)$, for all $\mathbf{y} \in \mathcal{Y}$;
  \item $\tilde{g}_{\ell}\left(\cdot, \mathbf{y}\right)$ is convex on $\mathcal{X}$ for all $\mathbf{y}\in \mathcal{Y}$, and $\tilde{g}_{\ell}\left(\mathbf{y}, \mathbf{y}\right) = g_{\ell}(\mathbf{y})$, for all $\mathbf{y}\in \mathcal{Y}$;
  \item $g_{\ell}(\mathbf{x}) \leq \tilde{g}_{\ell}\left(\mathbf{x}, \mathbf{y}\right)$ for all $\mathbf{x} \in \mathcal{X}$ and $\mathbf{y}\in \mathcal{Y}$;
  \item $\tilde{g}_{\ell}\left(\cdot, \cdot\right)$ is continuous on $\mathcal{X} \times \mathcal{Y}$ and $\nabla_{\mathbf{y}} \tilde{g}_{\ell}\left(\mathbf{y}, \mathbf{y}\right) = \nabla_{\mathbf{y}} g_{\ell}\left(\mathbf{y}\right)$, for all $\mathbf{y} \in \mathcal{Y}$;
  \item All feasible points of problem (\ref{prob:weight-sumrate-SU}) are regular (see, e.g. \cite{Scutari2014distributed}).
\end{enumerate}
where $\nabla_{\mathbf{y}} \tilde{f}\left(\mathbf{y}, \mathbf{y}\right)$ and $\nabla_{\mathbf{y}} \tilde{g}_{\ell}\left(\mathbf{y}, \mathbf{y}\right)$ denote the (partial) gradients of $\tilde{f}$ and $\tilde{g}_{\ell}$ respectively, which are with respect to the first argument evaluated at $\mathbf{y}$ (the second argument is kept fixed at $\mathbf{y}$).

Based on the above conditions, it is shown in~\cite[Theorem 2]{Scutari2014distributed} that the SCA algorithm converges to a stationary point of the noncovex problem (\ref{prob:weight-sumrate-SU}). Therefore, we conclude that each of the limit points generated by the proposed WMMSE-SCA algorithm is also a stationary point of problem (\ref{prob:weight-sumrate-SU}), which completes the proof of Theorem \ref{thm:SCAconvergence}.

\bibliographystyle{IEEEtran}
\bibliography{IEEEabrv,yuhanthesis}

\begin{IEEEbiography}[{\includegraphics[width=1in,height=1.25in,clip,keepaspectratio]{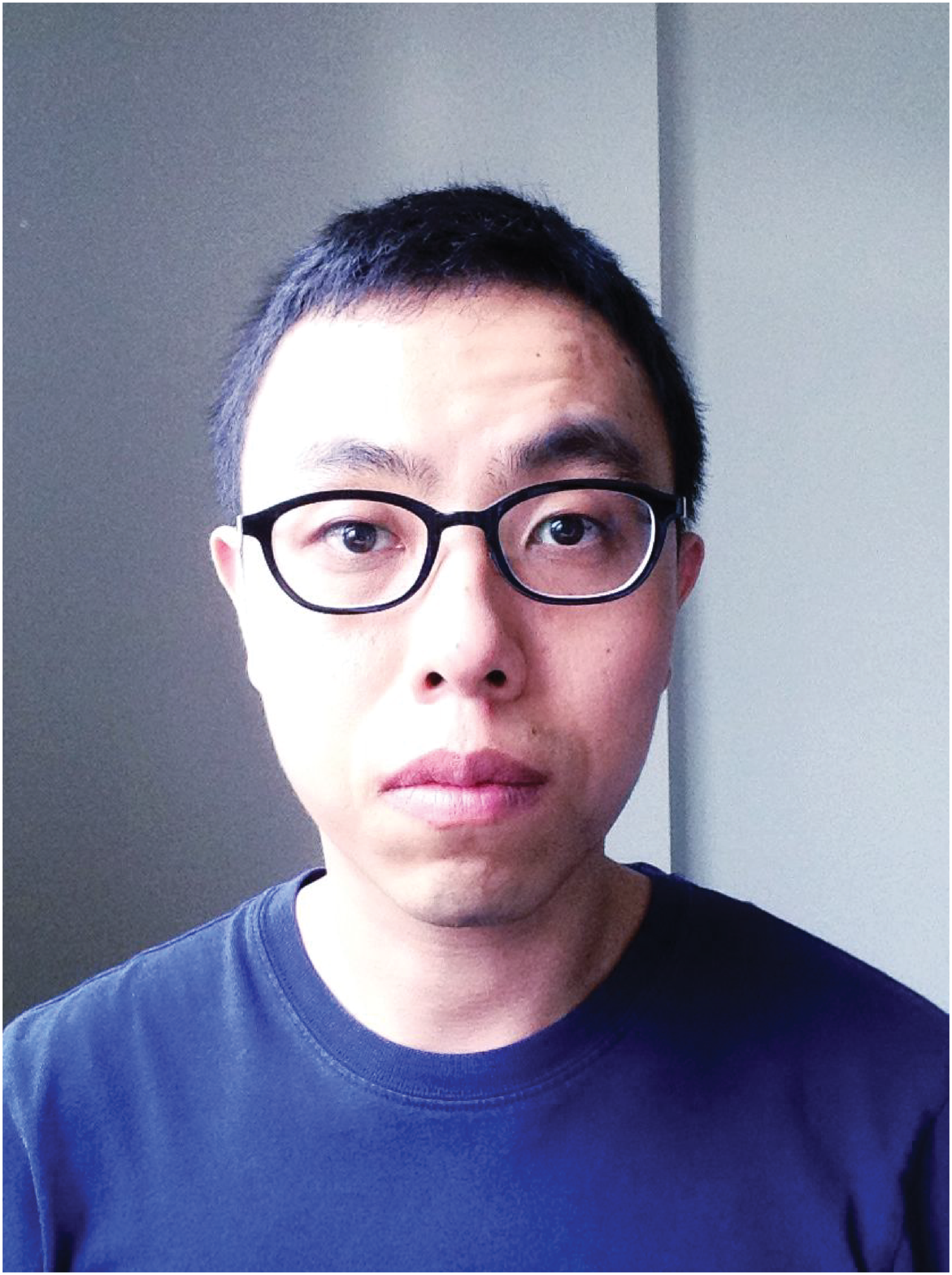}}]{Yuhan~Zhou}(S'08) received the B.E. degree in Electronic and Information Engineering from Jilin University, Jilin, China, in 2005, the M.A.Sc. degree from the University of Waterloo, ON, Canada, in 2009, and the Ph.D. degree from the University of Toronto, ON, Canada, in 2016, both in Electrical and Computer Engineering. Since 2016, he has been with Qualcomm Technologies Inc., San Diego, CA, USA. His research interests include wireless communications, network information theory, and convex optimization.
\end{IEEEbiography}
\begin{IEEEbiography}[{\includegraphics[width=1in,height=1.25in,clip,keepaspectratio]{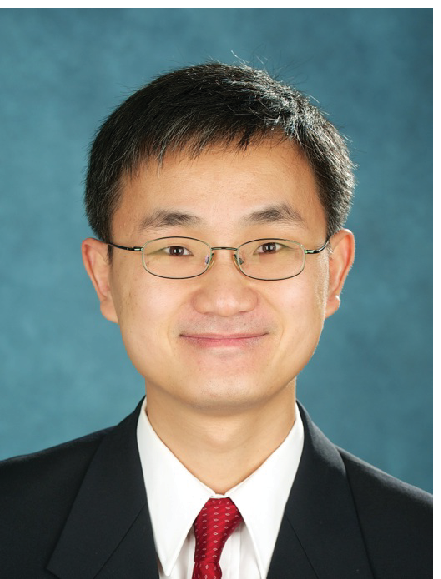}}]{Wei~Yu}(S'97-M'02-SM'08-F'14) received the B.A.Sc. degree in Computer Engineering and Mathematics from the University of Waterloo, Waterloo, Ontario, Canada in 1997 and M.S. and Ph.D. degrees in Electrical Engineering from Stanford University, Stanford, CA, in 1998 and 2002, respectively. Since 2002, he has been with the Electrical and Computer Engineering Department at the University of Toronto, Toronto, Ontario, Canada, where he is now Professor and holds a Canada Research Chair (Tier 1) in Information Theory and Wireless Communications. His main research interests include information theory, optimization, wireless communications and broadband access networks.

Prof. Wei Yu currently serves on the IEEE Information Theory Society
Board of Governors (2015-17). He is an IEEE Communications Society
Distinguished Lecturer (2015-16). He served as an Associate Editor for {\sc IEEE Transactions on Information Theory} (2010-2013), as an Editor for {\sc IEEE Transactions on Communications} (2009-2011), as an Editor for {\sc IEEE
Transactions on Wireless Communications} (2004-2007), and
as a Guest Editor for a number of special issues for the {\sc IEEE Journal on Selected Areas in
Communications} and the {\sc EURASIP Journal on Applied Signal Processing}. He was a Technical Program co-chair
of the IEEE Communication Theory Workshop in 2014, and a Technical
Program Committee co-chair of the Communication Theory Symposium at
the IEEE International Conference on Communications (ICC) in 2012. He
was a member of the Signal Processing for Communications and Networking
Technical Committee of the IEEE Signal Processing Society (2008-2013).
Prof. Wei Yu received a Steacie Memorial Fellowship in 2015, an IEEE
Communications Society Best Tutorial Paper Award in 2015, an IEEE ICC
Best Paper Award in 2013, an IEEE Signal Processing Society Best Paper
Award in 2008, the McCharles Prize for Early Career Research Distinction in
2008, the Early Career Teaching Award from the Faculty of Applied Science
and Engineering, University of Toronto in 2007, and an Early Researcher
Award from Ontario in 2006. He was named a Highly Cited Researcher by
Thomson Reuters in 2014.
\end{IEEEbiography}


\end{document}